\documentclass{iopart}

\usepackage{amssymb}
\usepackage{epsfig}
\usepackage{amscd}
\usepackage{graphicx,xspace}
\usepackage{pstricks}
\usepackage[left=3cm,top=3cm,right=3cm,bottom=3cm]{geometry}
\usepackage{color}

\begin{document}

\newcommand{\atanh}
{\operatorname{atanh}}
\newcommand{\ArcTan}
{\operatorname{ArcTan}}
\newcommand{\ArcCoth}
{\operatorname{ArcCoth}}
\newcommand{\Erf}
{\operatorname{Erf}}
\newcommand{\Erfi}
{\operatorname{Erfi}}
\newcommand{\Ei}
{\operatorname{Ei}}
\newcommand{\Pcal}
{{\mathcal P}}
\newcommand{\Prm}
{{\mathrm P}}

\title[Area distribution and the average shape of a L\'evy bridge]{Area distribution and the average shape of a L\'evy bridge}

\author{Gr{\'e}gory Schehr}

\address{Laboratoire de Physique Th\'eorique, CNRS-UMR 8627, Universit\'e Paris-Sud, 91405 Orsay Cedex, France }

\author{Satya N. Majumdar}

\address{Laboratoire de Physique Th\'eorique et Mod\`eles
  Statistiques, CNRS-UMR 8626, Universit\'e Paris-Sud, B\^at. 100, 91405 Orsay Cedex, France}

\date{Received:  / Accepted:  / Published }

\begin{abstract}
We consider a one dimensional L\'evy bridge $x_B$ of length $n$ and index $0 < \alpha < 2$, {\it i.e.} a L\'evy random walk constrained to start and end at the origin after $n$ time steps, $x_B(0) = x_B(n)=0$. We compute the distribution $P_B(A,n)$ of the area $A = \sum_{m=1}^n x_B(m)$ under such a L\'evy bridge and show that, for large $n$, it has the scaling form $P_B(A,n) \sim n^{-1-1/\alpha} F_\alpha(A/n^{1+1/\alpha})$, with the asymptotic behavior $F_\alpha(Y) \sim Y^{-2(1+\alpha)}$ for large $Y$. For $\alpha=1$, we obtain an explicit expression of $F_1(Y)$ in terms of elementary functions.  We also compute the average profile $\langle \tilde x_B (m) \rangle$ at time $m$ of a L\'evy bridge with fixed area $A$. For large $n$ and large $m$ and $A$, one finds the scaling form $\langle \tilde x_B(m) \rangle = n^{1/\alpha} H_\alpha\left({m}/{n},{A}/{n^{1+1/\alpha}} \right)$, where at variance with Brownian bridge, $H_\alpha(X,Y)$ is a non trivial function of the rescaled time $m/n$ and rescaled area $Y = A/n^{1+1/\alpha}$. Our analytical results are verified by numerical simulations.
\end{abstract}

\maketitle

\section{Introduction}

\begin{figure}[ht]
\begin{center}
 \includegraphics [width = \linewidth]{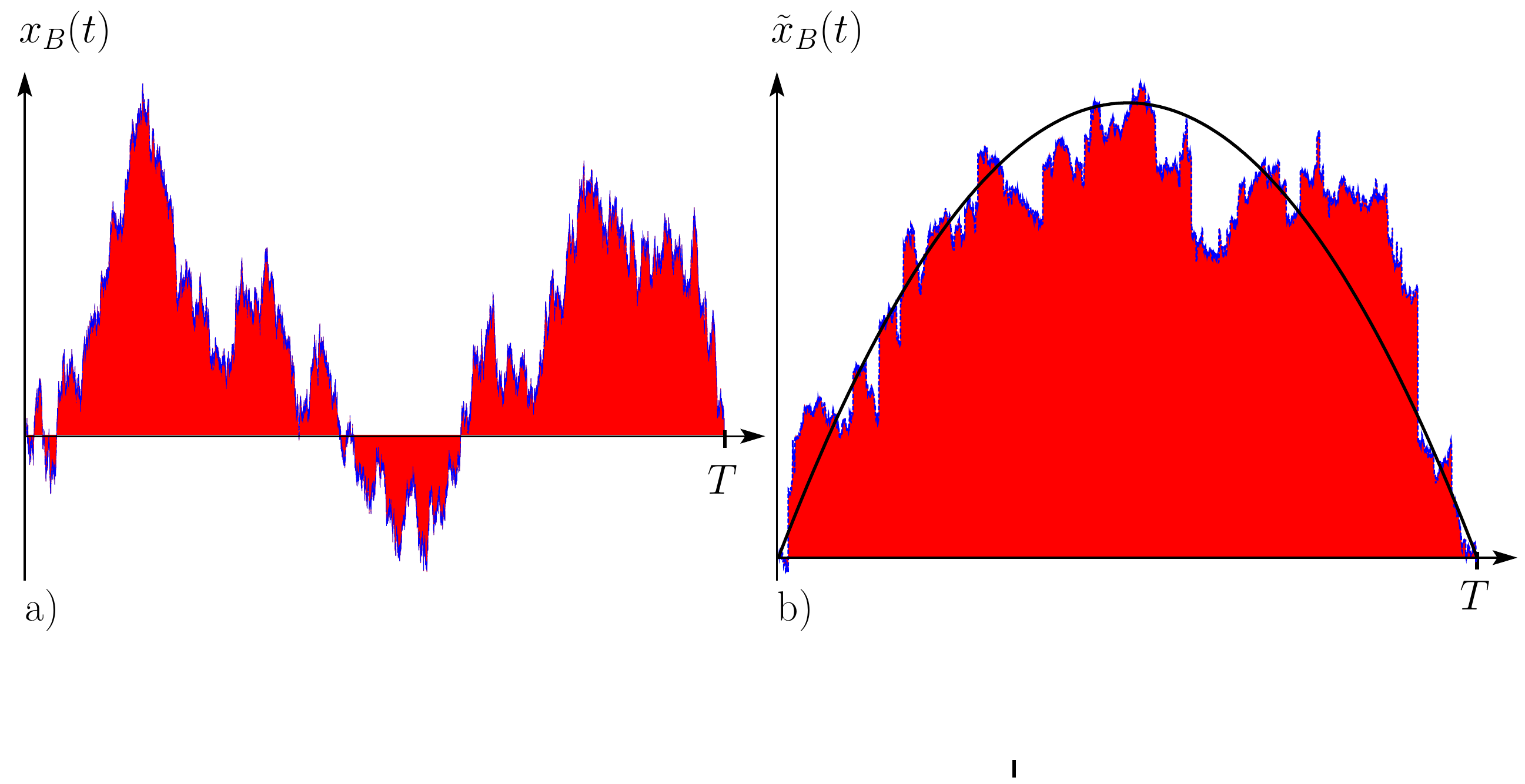}
 \end{center}
\caption{{\bf a)}: The blue line is the trajectory of a Brownian bridge $x_B(t)$ on $[0, T]$, $\it x_B(0)=x_B(T)=0$ and in red is the area $A$ under this Brownian bridge. {\bf b)}: The typical trajectory of a Brownian bridge $\tilde x_B(t)$ (in blue) with a fixed area $A$ (in red) on the interval $[0,T]$. The solid black line is the
average profile $\langle \tilde x_B(t) \rangle$ as given by Eq. (\ref{shape_bb}).}\label{fig_1}
\end{figure}

Random walks, and the associated continuous-time Brownian motion (BM),
are ubiquitous in nature. As such, they are not only
the cornerstones of statistical physics \cite{chandrasekhar, feller, hughes} but have also found many
applications in a variety of areas such as biology \cite{koshland},
computer science \cite{asmussen, satya_functionals}
and finance \cite{williams}. Continuous time Brownian motion is simply
defined by the equation of motion
\begin{eqnarray}\label{def_BM}
x(0) = 0 \;, \; \frac{\rmd x(t)}{\rmd t} = \eta(t) \;,
\end{eqnarray} 
where $\eta(t)$ is a Gaussian white noise of zero mean $\langle
\eta(t) \rangle = 0$ and short range correlations $\langle \eta(t)
\eta(t')\rangle = 2 D \delta(t-t')$ where $D$ is the diffusion
constant (in the following we set $D = 1$). An interesting variant
of Brownian motion in a given time interval $[0,T]$ is the so called
Brownian bridge $x_B(t)$ which is a Brownian motion conditioned to start and end at zero, {\it i.e.} $x_B(T) = x_B(0) = 0$. Here we
focus on two interesting observables associated with this bridge, namely
\begin{itemize}
\item{the distribution of the area $A$ under the bridge (see Fig. \ref{fig_1} a))
\begin{eqnarray}\label{def_A}
A = \int_0^T x_B(t)\, \rmd t \;,
\end{eqnarray}
which is obviously a random variable, being the sum of (strongly
correlated) random variables. For the Brownian bridge, the distribution of $A$ can easily be computed 
using the fact that $x_B(t)$ is a Gaussian random variable. This
can be seen from the well known identity in law \cite{feller}
\begin{eqnarray}\label{id_bb}
x_B(t) := x(t) - \frac{t}{T} x(T) \;,
\end{eqnarray}
where $x(t)$ is a standard Brownian motion (\ref{def_BM}). 
For the Brownian bridge, $A$ is thus also a centered Gaussian random variable. A direct
computation of the second moment $\langle A^2 \rangle$ yields
straightforwardly 
\begin{eqnarray}\label{dist_A_bb}
P_B(A,T) = \sqrt{\frac{3}{\pi T^3}} \exp{\left(-\frac{3 A^2}{T^3}
  \right)} \;. 
\end{eqnarray}
}
\item{the average shape of the bridge, $\langle \tilde x_B(t)\rangle$ for
  a fixed area $A$ (see Fig. \ref{fig_1} b)). For the Brownian bridge, it takes a simple form \cite{rivasseau}
\begin{eqnarray}\label{shape_bb}
\langle \tilde x_B(t)\rangle = \frac{A}{T} f\left(\frac{t}{T} \right)
\;, \; f(x) = 6x(1-x) \;.
\end{eqnarray}
} 

\end{itemize}

The distribution of the area under a Brownian bridge (\ref{dist_A_bb}) is a standard result and its extension to various constrained Brownian motions has recently 
attracted much attention \cite{satya_airy,kearney,schehr_airy,welinder,janson_review,rajabpour,rambeau_airy}. For instance, the distribution of the area under a Brownian excursion ({\it i.e.} a Brownian motion conditioned
to start and end at $0$ and constrained to stay positive in-between), the so called Airy-distribution, describes the statistics of 
the maximal relative height of one-dimensional elastic interfaces \cite{satya_airy, schehr_airy, rambeau_airy}. Another example is the area $A$
under a Brownian motion till its first-passage time $t_f$ \cite{kearney}, which has an interesting application to the description of the avalanches in  the directed Abelian sandpile model proposed in Ref. \cite{dhar}, such that $t_f$ relates to the avalanche duration and $A$ to the size of the avalanche cluster. Related quantities were recently studied in the statistics of avalanches near the depinning transition of elastic manifolds in random media \cite{pld}. On the other hand, the average shape of random walk bridges with a fixed area $A$ (\ref{shape_bb}) has been studied some time ago in the context of wetting \cite{rivasseau} to prove the validity of the Wulff construction in $1+1$ dimensions and more recently in the context of mass transport
models \cite{waclaw}. In these models, where the transport rules depend on the environment of the departure site, the steady state has a pair-factorized form \cite{satya_condensation}, which generalizes the factorized steady states found in simpler system like the zero range process \cite{evans_review, godreche_review,satya_review_condensation}. As the mass density crosses some critical
value, the system exhibits a condensation transition which is governed by interactions, which in turn give rise to a spatially extended condensate. It was shown in Ref. \cite{waclaw} that the shape of this condensate can be described by the average shape of a random walk bridge with fixed area, and the results
of Ref.~\cite{rivasseau} were recovered.

While these quantities are well understood for a Brownian bridge, much less is known for the case of a L\'evy bridge. The aim of the present paper is to compute the distribution of the area and the average shape for a fixed area $A$ in that case. To
this purpose, it is convenient to consider a random walk $x(m)$, in discrete time (see Fig. \ref{fig_rw} a)),
starting at $x(0)=x_0$ at time $0$ and evolving according to 
\begin{eqnarray}\label{rw}
x(m) = x(m-1) + \eta(m) \;, \
\end{eqnarray}
where $\eta(m)$ are independent and identically distributed (i.i.d.) random variables distributed according to a common distribution $\phi(\eta)$. Here we focus on the
case where $\phi(\eta) = {\cal S}_{\alpha}(\eta)$ where ${\cal S}_{\alpha}(\eta)$ is a symmetric $\alpha$-stable (L\'evy) distribution.
Its characteristic function is given by $\int_{-\infty}^\infty {\cal
  S}_\alpha(\eta) e^{i k \eta} \rmd \eta = e^{-  |k|^\alpha}$. In
particular, for large $\eta$, the distribution of $\eta$ has a power
law tail $\phi(\eta) \sim \eta^{-(1+\alpha)}$, with $0< \alpha < 2$. The
L\'evy bridge $x_B(m)$, on the interval $[0,n]$, is a L\'evy random
walk conditioned to start and end at $0$, {\it i.e.} $x_B(n)=x_B(0)=0$. In the following we will compute the distribution $P_B(A,n)$ of the area $A$ under the bridge of length $n$, {\it i.e.} $A = \sum_{m=0}^n x_B(m)$ and the average shape of a bridge $\langle \tilde x_B(m) \rangle$, with $0 \leq m \leq n$, for fixed $A$. Here we consider the natural scaling limit where $x_B \sim n^{1/\alpha}$, while $A \sim n^{1+1/\alpha}$, whereas the aforementioned previous works \cite{rivasseau, waclaw} focused on a different scaling limit which, for Brownian motion, corresponds to $x_B \sim \sqrt{n}$ and $A \sim n^2$. Note that, even in this natural scaling limit, the identity in law valid for the Brownian bridge (\ref{id_bb}) does
not hold for a L\'evy bridge \cite{knight, bertoin} and one thus expects the distribution of the area $A$ to be non trivial. Our results can be summarized as follows:
\begin{itemize}
 \item {for a free L\'evy random walk starting at $x_0 = 0$, one finds that the distribution of the area $P(A,n)$ takes the form

\vspace*{0.3cm}
\hspace*{4cm}\fbox{
\begin{minipage}[c]{0.5\textwidth}
\begin{eqnarray}
\hspace*{-2cm} P(A,n) = \frac{1}{\gamma_n} {\cal S}_{\alpha}\left(\frac{A}{\gamma_n} \right)\;, \; \gamma_n = \left(\sum_{m=1}^n m^\alpha \right)^{1/\alpha} \;, \nonumber
\end{eqnarray}
\end{minipage}
}\hfill
\begin{minipage}[c]{0.1\textwidth}
\centering
\begin{equation}
\hspace*{-10cm}
\end{equation}
\end{minipage}

\vspace*{0.3cm}
where ${\cal S}_\alpha(x)$ is a symmetric $\alpha$-stable distribution (\ref{def_stable}). For a L\'evy bridge, one finds, in the scaling limit $A \to \infty$, $n \to \infty$, keeping $A/n^{1+1/\alpha}$ fixed, that the distribution of the area $P_B(A,n)$ takes the scaling form

\hspace*{4cm}\fbox{
\begin{minipage}[c]{0.4\textwidth}
\begin{eqnarray}
\hspace*{-2cm} P_B(A,n) \sim \frac{1}{n^{1+1/\alpha}} F_{\alpha}\left(\frac{A}{n^{1+1/\alpha}} \right)\;, \nonumber
\end{eqnarray}
\end{minipage}
}\hfill
\begin{minipage}[c]{0.1\textwidth}
\centering
\begin{equation}
\hspace*{-10cm}
\end{equation}
\end{minipage}

\vspace*{0.3cm}
where $F_\alpha(y)$ is a monotonically decreasing function, with asymptotic behaviors
\vspace*{0.3cm}

\hspace*{4cm}\fbox{
\begin{minipage}[c]{0.4\textwidth}
\begin{eqnarray}
\hspace*{-2cm} F_\alpha(Y) \sim 
\cases{
 F_\alpha(0) \;, \; Y \to 0 \;, \\
 \frac{a_\alpha}{Y^{2(1+\alpha)}} \;, \; Y \to \infty \;,
}\nonumber
\end{eqnarray}
\end{minipage}
}\hfill
\begin{minipage}[c]{0.1\textwidth}
\centering
\begin{equation}
\hspace*{-10cm}
\end{equation}
\end{minipage}

\vspace*{0.3cm}
where $F_\alpha(0)$, see Eq. (\ref{expr_constant}), and $a_\alpha$, see Eq. (\ref{large_y}), are computable constants. For $\alpha=1$, we obtain an explicit expression for $F_1(Y)$ in terms of elementary functions (\ref{elem}).}
\item{on the other hand, in the aforementioned scaling limit, one
  obtains the average profile $\langle \tilde x_B(m) \rangle$ for a L\'evy
  bridge as well as the average profile $\langle \tilde x(m) \rangle$ for a
  free L\'evy walk with a fixed area $A$. For $\langle \tilde x(m) \rangle$ one obtains a farely simple expression

\vspace*{0.3cm}
\hspace*{4cm}\fbox{
\begin{minipage}[c]{0.4\textwidth}
\begin{eqnarray}
\hspace*{-2cm}  \langle \tilde x(m) \rangle = \frac{A}{n} \frac{\alpha+1}{\alpha} \left[1-\left(1-\frac{m}{n} \right)^\alpha \right] \;. \nonumber
\end{eqnarray}
\end{minipage}
}\hfill
\begin{minipage}[c]{0.1\textwidth}
\centering
\begin{equation}
\hspace*{-10cm}
\end{equation}
\end{minipage}

\vspace*{0.3cm}
For a L\'evy bridge, the expression is more involved. For generic $\alpha$ one finds the scaling form

\vspace*{0.3cm}
\hspace*{4cm}\fbox{
\begin{minipage}[c]{0.4\textwidth}
\begin{eqnarray}
\hspace*{-2cm}   \langle \tilde x_B(m) \rangle = n^{1/\alpha} H_\alpha\left(\frac{m}{n},\frac{A}{n^{1+1/\alpha}} \right)  \;, \nonumber
\end{eqnarray}
\end{minipage}
}\hfill
\begin{minipage}[c]{0.1\textwidth}
\centering
\begin{equation}
\hspace*{-10cm}
\end{equation}
\end{minipage}

\vspace*{0.3cm}
which, in general, has a non-trivial dependence on $A$. One recovers a linear dependence in $A$ (as for the Brownian bridge in Eq. (\ref{shape_bb})) only in the limits $A \to 0$ (\ref{small_y_shape}) and $A \to \infty$ (\ref{large_y_shape}). 

}

\end{itemize}

The paper is organized as follows. In section 2, we compute the joint distribution of the position and the area under a L\'evy random walk. In section 3, we use these results to compute the distribution of the area $A$ under a L\'evy bridge of size $n$ while in section 4, we use them to compute the average profile of a L\'evy walk 
with a fixed area. Finally, in section 5 we present a numerical method, based on a Monte-Carlo algorithm, to compute numerically $P_B(A,n)$ and $\langle \tilde x_B(m)\rangle$ before we conclude in section 6. Some technical (and useful) details have been left in Appendices A,B and C.

\section{Free L\'evy walk : joint distribution of the position and the area}

\begin{figure}
\centering
 \includegraphics[width = \linewidth]{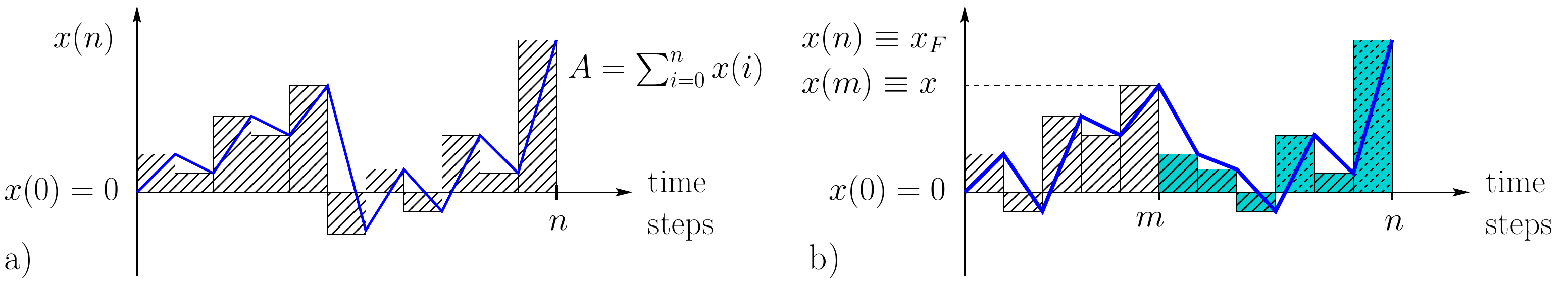}
\caption{{\bf a)}: The area $A$ under a random walk in discrete time. {\bf b)}: Illustration of the method, see Eq. (\ref{start_free}), to compute the probability $\tilde P(x,m|A,n)$ that the position of the random walker, starting in $x(0)=0$ is $x$ after $m$ time steps, given that the area, after $n$ time steps, is fixed to $A$. The light area corresponds to $A_1$, the area under the walk over the interval $[0,m]$ while the shaded area corresponds to the area $A-A_1$, the area under the walk over the interval $[m,n]$.}\label{fig_rw}
\end{figure}

We start with the computation of the joint distribution $P(x,A,m|x_0,x_0,0)$ of the position and the area after $m$ steps given that $x(0)=x_0$ (see Fig. \ref{fig_rw} a)). If we denote by $A(m)$ the area under the random walk after $m$ time steps, this random variable evolves according to the equation
\begin{eqnarray}\label{area}
&&A(0) = x_0  \;, \\
&&A(m) = A(m-1) + x(m) \;.
\end{eqnarray}
Therefore $P(x,A,m|x_0,x_0,0)$ satisfies the following recursion relation:
\begin{eqnarray}\label{recurrence}
 &&P(x,A,0|x_0,x_0,0) = \delta(x-x_0) \delta(A-x_0) \;, \nonumber \\
 &&P(x,A,m|x_0,x_0,0) = \int_{-\infty}^\infty P(x-\eta, A-x,m-1|x_0,x_0,0) \phi(\eta) \rmd \eta \;.
\end{eqnarray}
Introducing $\hat \phi(k) = \int_{-\infty}^\infty \phi(\eta) e^{i k \eta} \rmd \eta$ the Fourier transform of $\phi(\eta)$, and thus 
$\hat \phi(k) =  e^{-|k|^\alpha}$ for a L\'evy random walk, and $\hat P(k_1, k_2,m|x_0,x_0,0)$ the double Fourier transform of $P(x,A,m|x_0,x_0,0)$ 
with respect to both $x$ and $A$, {\it i.e.} $\hat P(k_1, k_2,m|x_0,x_0,0) = \int_{-\infty}^\infty \rmd x \int_{-\infty}^\infty \rmd A  P(x,A,m|x_0,x_0,0) e^{i k_1 x + i k_2 A}$ the
recursion relation (\ref{recurrence}) reads
\begin{eqnarray}\label{recurrence_fourier}
&& \hat P(k_1,k_2,0|x_0,x_0,0) = e^{i (k_1+k_2) x_0 } \;, \\
&& \hat P(k_1, k_2,m|x_0,x_0,0) = \hat \phi(k_1 + k_2) \hat P(k_1+k_2,k_2,m-1|x_0,x_0,0) \;,
\end{eqnarray}
which can be solved, yielding
\begin{eqnarray}\label{expr_Fourier_joint}
\hat P(k_1, k_2,n|x_0,x_0,0) = \prod_{m=1}^n \hat \phi(k_1 + m k_2) e^{i (k_1 + (n+1)k_2) x_0} \;.
\end{eqnarray}
Hence for a L\'evy walk of index $\alpha$ one has simply
\begin{eqnarray}\label{start_expr}
\fl P(x,A,n|x_0,x_0,0) = \int_{-\infty}^\infty \frac{\rmd k_1}{2 \pi} \int_{-\infty}^\infty \frac{\rmd k_2}{2 \pi} e^{- \sum_{m=1}^n |k_1 + m k_2|^\alpha}e^{- ik_1 (x-x_0)} e^{-ik_2 (A-(n+1)x_0)} \;.
\end{eqnarray}
Note that this expression (\ref{start_expr}) can also be obtained directly by noticing that $x(n) = x_0 + \sum_{m=1}^n \eta(i)$ and thus $A(n) - (n+1)x_0 = \sum_{m=1}^n x(m) = \sum_{m=1}^n \sum_{l=1}^m \eta(l) = \sum_{m=1}^n m \eta(n+1-m)$ such that
\begin{eqnarray}\label{direct_joint}
\fl && P(x,A,n|x_0,x_0,0) = \prod_{m=1}^n \int_{-\infty}^\infty \rmd
\eta(m)  \prod_{m=1}^n \phi\left[\eta(m) \right]\delta\left(x-x_0-\sum_{m=1}^n
\eta(m)\right) \nonumber \\ 
&& \times \delta\left(A - (n+1)x_0 - \sum_{m=1}^n m \eta(n+1-m)\right) \;.
\end{eqnarray}
After a double Fourier transform with respect to $x$ and $A$, this
Eq. (\ref{direct_joint}) yields immediately the expression $\hat
P(k_1, k_2,n|x_0,x_0,0)$ in Eq. (\ref{expr_Fourier_joint}). Of course
the marginal distribution of the position $P(x,n|x_0,0)$ and of
the area $P(A,n|x_0,0)$ are also stable laws. Indeed one has
\begin{eqnarray}\label{marginals}
\fl && P(x,n|x_0,0) = \int_{-\infty}^\infty P(x,A,n|x_0,x_0,0) {\rm d} A = \frac{1}{n^{1/\alpha}} {\cal S}_\alpha \left( \frac{x-x_0}{n^{1/\alpha}} \right) \;, \\
\fl && P(A,n|x_0,0) = \int_{-\infty}^\infty P(x,A,n|x_0,x_0,0) {\rm d} x = \frac{1}{\gamma_n} {\cal S}_\alpha \left( \frac{A-(n+1)x_0}{\gamma_n} \right) \;, \nonumber \\
\fl && \hspace*{1.6cm}\gamma_n = \left( \sum_{m=1}^n m^\alpha \right)^{1/\alpha} \sim \frac{n^{1+1/\alpha}}{(\alpha+1)^{1/\alpha}} \;, \; n \gg 1 \;, \nonumber
\end{eqnarray}
where 
\begin{eqnarray}\label{def_stable}
 {\cal S}_\alpha(x) = \int_{-\infty}^\infty \, e^{-|k|^\alpha - i k x} \frac{\rmd k}{2 \pi} \;.
\end{eqnarray}
For example, ${\cal S}_1(x)$ is the Cauchy distribution while ${\cal S}_2(x)$ is a Gaussian distribution :
\begin{eqnarray}\label{stable_explicit}
{\cal S}_1(x) = \frac{1}{\pi} \frac{1}{1+x^2} \;, \; {\cal S}_2(x) = \frac{1}{2 \sqrt{\pi}} e^{-\frac{x^2}{4}} \;.
\end{eqnarray}
Note also the explicit expression  
\begin{eqnarray}\label{salpha_zero}
 {\cal S}_\alpha(0) = \int_{-\infty}^\infty \, e^{-|k|^\alpha} \frac{\rmd k}{2 \pi} = \frac{\Gamma(1+\alpha^{-1})}{\pi} \;,
\end{eqnarray}
which will be useful in the following. 

We now want to study $P(x,A,n|x_0,x_0,0)$ in the limit of large $n$. The marginal distributions in Eq. (\ref{marginals}) suggest the scaling $x \sim n^{1/\alpha}$ and $A \sim n^{1/\alpha + 1}$. From the expression in Eq. (\ref{start_expr}) one checks explicitly that in the limit $n \to \infty$, keeping $X = x/n^{1/\alpha}$ and $Y = A/n^{1/\alpha+1}$ fixed, the joint distribution takes the scaling form
\begin{eqnarray}\label{scaling_form}
 P(x,A,n|x_0,x_0,0) = \frac{1}{n^{2/\alpha+1}} G\left(
 \frac{x}{n^{1/\alpha}}, \frac{A}{n^{1/\alpha+1}} \bigg |
 \frac{x_0}{n^{1/\alpha}} \right) \;,
\end{eqnarray}
where the function $G(X,Y|X_0)$ is given by
\begin{eqnarray} 
\fl G(X,Y|X_0) &=&  \int_{-\infty}^\infty \frac{\rmd k_1}{2 \pi} \int_{-\infty}^\infty \frac{\rmd k_2}{2 \pi} e^{-\int_0^1 |k_1 + k_2 z |^\alpha dz - i k_1 (X-X_0) - i k_2 (Y-X_0) } \;.
\end{eqnarray}
After the change of variable $k_2 = k$ and $k_1 = k r$, we obtain
\begin{eqnarray} \label{expr_scaling}
\fl G(X,Y|X_0) &=& \int_{-\infty}^\infty \frac{\rmd r}{2\pi} \int_{-\infty}^\infty\frac{\rmd k}{2\pi} |k| e^{-|k|^\alpha \gamma(r) - i  k r (X-X_0 ) - i k (Y - X_0)}
 \;, \; \gamma(r) = \int_0^1 |r+z|^\alpha \rmd z \;,
\end{eqnarray}
where the function $\gamma(r)$ is explicitly given by
\begin{eqnarray}\label{expr_gamma}
 \gamma(r) = 
\cases{
 \frac{1}{\alpha+1} \left( (-r)^{\alpha+1} - (-1-r)^{\alpha+1} \right) \;, \; r < -1 \;, \\ 
\frac{1}{\alpha+1} \left((r+1)^{\alpha+1} + (-r)^{\alpha+1} \right) \;, \; - 1 \leq r \leq 0 \;, \\
 \frac{1}{\alpha+1}\left((r+1)^{\alpha+1} - r^{\alpha+1} \right) \;, \; r > 0 \;.
}
\end{eqnarray}
For $\alpha=2$, one has $\gamma(r) = r^2+r+1/3$ and one finds
\begin{eqnarray}
 G(X,Y|X_0) = \frac{\sqrt{3}}{2 \pi} \exp{\left[- 3(Y-X_0)(Y-X) - (X-X_0)^2 \right]} \;,
\end{eqnarray}
which yields back the propagator of the so-called random acceleration process \cite{rap}.

\section{L\'evy bridge}

In the absence of any constraint for the walker, the area under a L\'evy random walk is the sum of L\'evy random variables
and is thus again a L\'evy random variable. However, if one considers constrained L\'evy walks, this is not true anymore 
and the area may become different from a simple L\'evy random variable. In the first subsection, we compute the distribution of the position for a L\'evy bridge while the second subsection is devoted to the distribution of the area under this bridge.

\subsection{Distribution of the position}

Here we study the L\'evy bridge $\{ x_B(m) \}_{0\leq m \leq n}$ which starts at $0$ at time $0$, $x_B(0) =0$, and is constrained to come back to $0$ after $n$ time steps, {\it i.e.} $x_B(n) =0$. In that case, one can compute the distribution of the position $P_B(x,m)$ after $m$ time steps for such a bridge as  $P_B(x,m) = P(x,m|0,0) P(x,n-m|0,0)/P(0,n|0,0)$, such that in the scaling limit one has
\begin{eqnarray}\label{marginal_bridge1}
 && P_B(x,m) = \frac{1}{n^{1/\alpha}} G_B\left(\frac{x}{n^{1/\alpha}},\frac{m}{n} \right) \;, \; \nonumber \\
&& G_B(X,\tau) = \frac{\pi}{\Gamma(1+\alpha^{-1})}\frac{1}{(\tau(1-\tau))^{1/\alpha}} {\cal S}_\alpha \left(\frac{X}{\tau^{1/\alpha}} \right) {\cal S}_\alpha \left( \frac{X}{(1-\tau)^{1/\alpha}}\right) \;,
\end{eqnarray}
where we have used ${\cal S}_\alpha(0) = \Gamma(1+\alpha^{-1})/\pi$, see Eq. (\ref{salpha_zero}). For $\alpha=2$ it is easy to see from Eq. (\ref{marginal_bridge1}) that $x_B(m)$ is a Gaussian variable. However, for $\alpha < 2$, the L\'evy bridge is not any more a L\'evy random variable. For instance, for $\alpha = 1$ one obtains a non trivial distribution
\begin{eqnarray}\label{marginal_bridge2}
 G_B(X,\tau) = \frac{1}{\pi} \frac{\tau(1-\tau)}{(\tau^2 + X^2)((1-\tau)^2+X^2)} \;.
\end{eqnarray}
It is also easy to see that, for any $\alpha < 2$ one has the asymptotic behavior
\begin{eqnarray}\label{asympt_bridge}
 G_B(X,\tau) \sim c'_\alpha \tau(1-\tau) X^{-2(\alpha+1)} \;, \; X \gg 1 \;,
\end{eqnarray}
where $c'_\alpha$ is independent of $\tau$, which implies that $\langle (x_B(m))^2 \rangle$ is well defined for $\alpha > 1/2$ (of course $\langle x_B(m) \rangle = 0$ by symmetry for all $\alpha$). A straightforward calculation shows that
\begin{eqnarray}\label{variance_bridge}
\frac{\langle x^2_B(m) \rangle}{n^{2/\alpha}} =  \tilde a_\alpha \frac{m}{n}\left(1-\frac{m}{n} \right) \;, \; \tilde a_\alpha = \frac{\alpha \Gamma(2-\alpha^{-1})}{\Gamma(1+\alpha^{-1})} \;.
\end{eqnarray}
It is interesting to notice that ${\langle x^2_B(m) \rangle}/{n^{2/\alpha}}$ depends on $\alpha$ only through the amplitude $\tilde a_\alpha$ but the parabolic shape in $(m/n)(1-m/n)$ holds for all values of $2 \geq \alpha > 1/2$. Besides $\tilde a_\alpha$ is diverging for $\alpha \to 1/2^+$ while one has ${\tilde a}(1) = 1$ and ${\tilde a}(2) = 2$. One finds, curiously, that it reaches a minimum for a non-trivial value $\alpha^* = 0.74122 \dots$ for which ${\tilde a}(\alpha^*) = 0.85264 \dots$. In view of these properties (\ref{marginal_bridge1}, \ref{marginal_bridge2}) one expects that, for $\alpha < 2$, the area under such a L\'evy bridge has a non trivial distribution, which we now focus on. 

\subsection{Distribution of the area}

In this subsection, we consider a L\'evy bridge, {\it i.e.} a L\'evy walk which starts at the origin $x_0=0$ and is conditioned to come back to the origin
after $n$ steps and we ask : what is the distribution ${P}_B(A,n)$ of the area $A$ under this {\it L\'evy} bridge ? One 
can obtain ${P}_B(A,n)$ from Eqs (\ref{start_expr}, \ref{marginals}) as
\begin{eqnarray}
 {P}_B (A,n) = \frac{P(0,A,n|0,0,0)}{P(x=0,n|0,0)} \;.
\end{eqnarray}
Therefore, using $P(x=0,n|0,0) = n^{-1/\alpha} {\cal S}_\alpha(0) = n^{-1/\alpha} \Gamma(1+\alpha^{-1})/\pi$, see Eq. (\ref{salpha_zero}), together with the scaling form (\ref{scaling_form}, \ref{expr_scaling}) one obtains, in the limit $n \to \infty$, keeping $A/n^{1+1/\alpha}$ fixed:  
\begin{eqnarray}\label{expr_scaling_genalpha}
 {P}_B (A,n) = \frac{1}{n^{1+1/\alpha}}
 {F}_\alpha\left(\frac{A}{n^{1+1/\alpha}} \right) \;, \nonumber \\
{F}_\alpha(Y) =\frac{1}{2\Gamma(1+\alpha^{-1})} \int_{-\infty}^\infty \, \rmd r \int_{-\infty}^\infty \frac{\rmd k}{2 \pi} |k| e^{-|k|^\alpha \gamma(r) - i Y k} \;.
\end{eqnarray}
Using the explicit expression of $\gamma(r)$ above (\ref{expr_gamma}), one computes the Fourier transform $\hat {F}_\alpha(k)$ of ${F}_\alpha(Y)$ as
\begin{eqnarray}\label{expr_fourier_gen_alpha}
&&\hat {F}_\alpha(k) = \int_{-\infty}^\infty  {F}_\alpha(Y) e^{i k Y} \rmd Y \;, \\
 && = \frac{|k|}{\Gamma(1+\frac{1}{\alpha})} \left[ \int_0^\infty e^{-\frac{|k|^\alpha}{\alpha+1} \left[(r+1)^{\alpha+1}-r^{\alpha+1}\right]}  \rmd r + \int_0^{1/2}    e^{-\frac{|k|^\alpha}{\alpha+1} \left[(1/2+r)^{\alpha+1}+(1/2-r)^{\alpha+1}\right]} \rmd r 
\right] \;. \nonumber 
\end{eqnarray}
For generic $\alpha$, it seems quite difficult to perform explicitly the integrals over $r$ and $k$  in the expression for the distribution ${F}_\alpha(Y)$ in Eq. (\ref{expr_scaling_genalpha}). One can however extract from this expression the asymptotic behaviors both for $Y \to 0$ and $Y \to \infty$. 

{\bf Asymptotic behavior for small argument.} For small argument, it is straightforward to see on the expression (\ref{expr_scaling_genalpha}) above that the leading behavior of $F_\alpha(Y)$ when $Y \to 0$ is given by
\begin{eqnarray}\label{expr_constant}
F_\alpha(Y) \sim F_\alpha(0) \;, \; Y \to 0 \;, {\rm with} \; \nonumber \\
\fl F_\alpha(0) = \frac{(\alpha+1)^{\frac{2}{\alpha}}}{2 \pi} \frac{\Gamma(1+\frac{2}{\alpha})}{\Gamma(1+\frac{1}{\alpha})} \left( \int_0^\infty \frac{\rmd r}{((r+1)^{\alpha+1} - r^{\alpha+1})^{\frac{2}{\alpha}}} + \int_0^{1/2} \frac{\rmd r}{((1/2+r)^{\alpha+1} + (1/2-r)^{\alpha+1} )^{\frac{2}{\alpha}}}\right) \nonumber \;.
\end{eqnarray}
A study of this function $F_\alpha(0)$ shows that it is a decreasing function of $\alpha$ on the interval $]0,2]$, which is diverging when $\alpha \to 0$. For $\alpha = 1$ and $\alpha =2 $, $F_\alpha(0)$ assumes simple values
\begin{eqnarray}
 F_1(0) = 1 + \frac{4}{\pi} = 2.27324... \;, \;  F_2(0) = \sqrt{\frac{3}{\pi}} = 0.977205...
\end{eqnarray}

{\bf Asymptotic behavior for large argument.} The analysis of the large argument behavior of $F_\alpha(Y)$ is more involved. A careful analysis, left in \ref{app_asympt}, shows that  for large $Y$ one has:
\begin{eqnarray}\label{large_y}
&& F_\alpha(Y) \sim \frac{a_\alpha}{Y^{2(1+\alpha)}} \;, \; Y \gg 1 \;, \\
&& a_\alpha = \frac{2^{-2(2+\alpha)} \sqrt{\pi} \Gamma(2+2\alpha) \tan{(\alpha \pi/2)}}{\Gamma(2+\alpha^{-1}) \Gamma(1-\alpha) \Gamma(\frac{5}{2} + \alpha)} \;.
\end{eqnarray}
When $\alpha \to 2$, one has from (\ref{large_y}), $a_{\alpha} \sim \frac{\sqrt{\pi}}{21}(\alpha-2)^2$. From Eq. (\ref{large_y}) one obtains that the second moment of the distribution $\langle Y^2 \rangle$ is defined only for $\alpha > 1/2$ where it takes the value (see Eq. (\ref{expr_appendix})) 
\begin{eqnarray}\label{area_variance}
 \langle Y^2 \rangle = \frac{{\tilde a}_\alpha}{12} = \frac{\alpha \Gamma(2-\alpha^{-1})}{12 \Gamma(1+\alpha^{-1})} \;, \; \alpha > 1/2 \;,
\end{eqnarray}
where the amplitude ${\tilde a}_\alpha$ appears in the expression for $\langle x^2_B(m) \rangle$ computed above (\ref{variance_bridge}). This power law tail of the area distribution (\ref{large_y}) with an exponent $2(1+\alpha)$ is quite interesting. Indeed, the area itself is the sum of non-identical and strongly correlated variables $x_B(m)$ all having a similar power law tail also with exponent $2(1+\alpha)$ (\ref{asympt_bridge}). For $\alpha > 1/2$, their variance is finite and the non-Gaussianity
of $A$ can a priori be due both to the correlations between the $x_B(m)$'s and to the fact that $A$ is the sum of non-identical random variables. To test which of these features is responsible for the non-Gaussianity of $A$, we study the sum of $n$ random variables $X_m$ which are {\it independent} and such that $X_m$ has the same distribution as $x_B(m)$. Defining $S_n = \sum_{m=1}^n X_m$ and $\Sigma_n^2 = \sum_{m=1}^n \langle X_m^2 \rangle$, it is known that $S_n/\Sigma_n$ converges to a centered Gaussian variable of unit variance if the following condition (known as the Lindeberg's condition) is satisfied \cite{feller}
\begin{eqnarray}\label{lindeberg}
 \lim_{n \to \infty} \frac{1}{\Sigma_n^2} \int_{|x| > \epsilon \Sigma_n} x^2 {\rm Proba}(X_m = x) \rmd x = 0 \;, \; \forall \epsilon > 0 \;. 
\end{eqnarray}
Intuitively, this Lindeberg condition (\ref{lindeberg}) ensures that the probability that any term $X_m$ will be of the same order of magnitude as the sum $S_n$ must tend to zero (see \ref{appendix_lindeberg} for an example of non-identical independent random variables which do not satisfy the Lindeberg condition). In the present case of the L\'evy bridge, one can check (see \ref{appendix_lindeberg}) that if $X_m$ is distributed like $x_B(m)$ then the above Lindeberg condition (\ref{lindeberg}) is satisfied (see \ref{appendix_lindeberg}). Therefore the deviations from Gaussianity (\ref{large_y}) are purely due to the {\it strong correlations} between the positions of the walker $x_B(m)$'s.

{\bf The special case $\alpha = 1$}. In the Cauchy case, $\alpha=1$, the integral over $k$ can be done in Eq. (\ref{expr_scaling}) to obtain
\begin{eqnarray}
 G(X,Y|0) = \int_{-\infty}^\infty \frac{\gamma^2(r) - (r X + Y)^2}{\left(\gamma^2(r) + (r X + Y)^2\right)^2} \frac{\rmd r}{2 \pi^2}  \;,
\end{eqnarray}
where the function $\gamma(r)$ (\ref{expr_gamma}) takes here a rather simple form
\begin{eqnarray}
  \gamma(r) = 
\cases{ - r - \frac{1}{2} \;, \; r < -1 \\ 
r^2+r + \frac{1}{2} \;, \; - 1 \leq r \leq 0 \\
r + \frac{1}{2}  \;, r > 0 \;.
}
\end{eqnarray}
The distribution of the area under a Cauchy bridge is thus given by 
 \begin{eqnarray}\label{cauchy_integral}
&& {P}_B (A,n) = \frac{1}{n^{2}} F_1 \left(\frac{A}{n^{2}} \right) \;, \nonumber \\
&& F_1 (Y) = \frac{1}{\pi}\frac{2}{1+4Y^2} + \frac{1}{\pi} \int_0^{1/2} \frac{(u^2+1/4)^2 - Y^2}{((u^2+1/4)^2+Y^2)^2} \rmd u \;.
\end{eqnarray}
Under this form (\ref{cauchy_integral}), one can easily obtain the asymptotic behaviors as
\begin{eqnarray}
 F_1(Y) \sim
\cases{
1 + \frac{4}{\pi} \;, \; Y \to 0 \\
\frac{1}{20 \pi Y^4} \;, \; Y \to \infty
}
\end{eqnarray}
 in agreement with the asymptotic behaviors obtained above (\ref{expr_constant}, \ref{large_y}). In fact the integral over $u$ in the expression
above (\ref{cauchy_integral}) can be done explicitly yielding the expression
\begin{eqnarray}\label{expr_arctan}
&& F_1(Y) = \frac{1}{\pi}\frac{2}{1+4Y^2} \\
 &&+ \frac{1}{\pi} \left( \frac{2(1-8Y^2)}{(1+4Y^2)(1+16Y^2)} + \frac{4}{(1+16Y^2)^{\frac{3}{2}}} {\rm Re}\left[{(1- 4 i Y)^{\frac{3}{2}} \arctan{\left((1+4 i Y)^{-\frac{1}{2}}\right)}}\right] \right) \;, \nonumber
\end{eqnarray}
where ${\rm Re}{(z)}$ denotes the real part of the complex number $z$. In \ref{appendix_elementary} we show how this expression~(\ref{expr_arctan})
can be written explicitly in terms of elementary functions (\ref{def_ab}, \ref{def_lm}, \ref{elem}).

\section{Average profile for fixed area $A$.}

\subsection{The case of a free L\'evy walk}

\begin{figure}[ht]
\begin{center}
 \includegraphics [width = \linewidth]{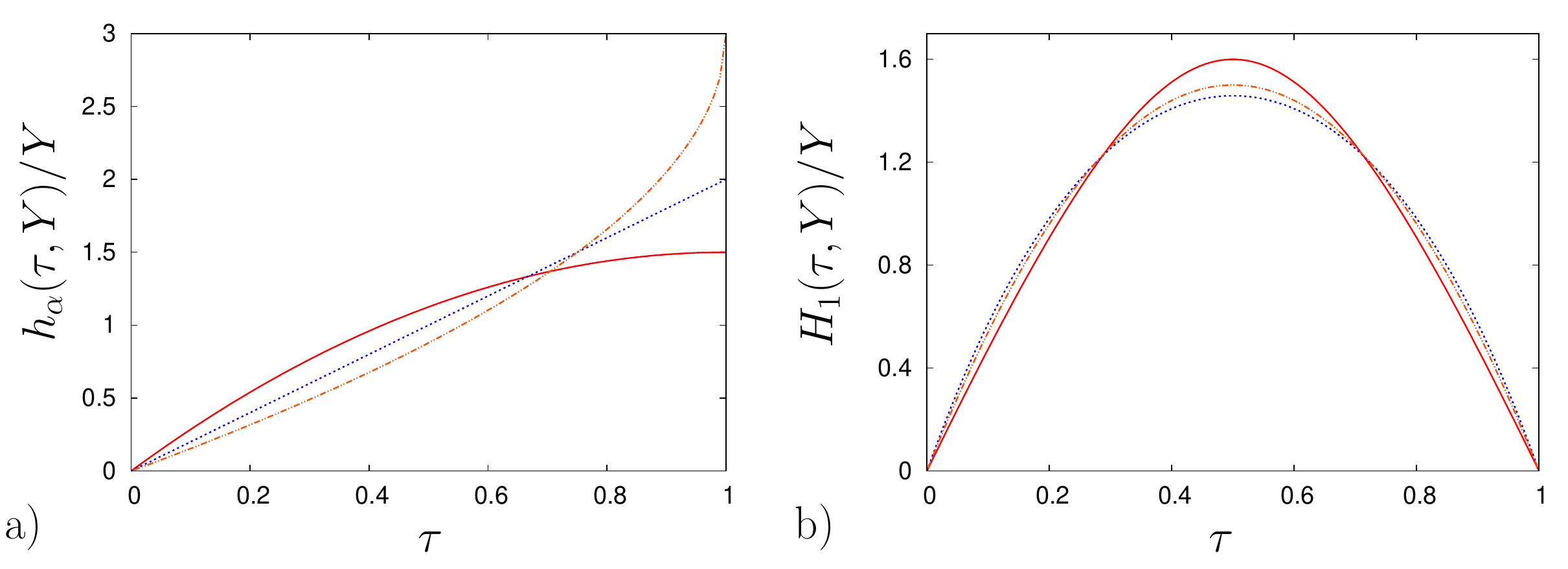}
 \end{center}
\caption{Average shape of a free L\'evy walk in {\bf a)} and of a L\'evy bridge in {\bf b)}. {\bf a)}: Plot of $h_{\alpha}(\tau,Y)/Y$ given in Eq. (\ref{explicit_free}) as a function of $\tau$ for $\alpha = 2$ (solid line), $\alpha=1$ (dotted line) and $\alpha=1/2$ (dashed line). {\bf b)}: Plot of $H_1(\tau,Y)/Y$ as a function of $\tau$ in the limit $A \to 0$ (solid line), as given by Eq. (\ref{small_y_shape}), and in the limit $A \to \infty$ (dotted line), as given in Eq. (\ref{large_y_shape}). For comparison, we have also plotted the parabola (dashed line) corresponding to the result for a Brownian bridge, {\it i.e.} $\alpha=2$, as given in Eq. (\ref{shape_bb})  }\label{fig_profiles}
\end{figure}

We first consider the case of a free L\'evy walk of constrained
area. We compute the probability $\tilde P(x,m | A,n)$ that the
position of the random walker, starting at $x(0)=x_0=0$ at time $0$, is $x$
after $m$ time steps given that 
the area, after $n$ time steps, is fixed to $A$. From this
probability, one obtains the average profile as $\langle \tilde x(m)
\rangle = \int_{-\infty}^\infty x \tilde P(x,m|A,n) \rmd x$. To compute this 
probability $\tilde P(x,m|A,n)$, we divide the interval $[0,n]$ into two intervals $[0,m]$ and $[m,n]$. Over
$[0,m]$ the process starts at $x_0 = 0$ with area $A_0=0$ and reaches to $x$ with area $A_1$ (see the light area on Fig. \ref{fig_rw} b)). Over the interval 
$[m,n]$, the process starts in $x$ and reaches to $x_F$ with area $A-A_1$ (see the shaded area on Fig. \ref{fig_rw} b)). Therefore this probability $\tilde P(x,m|A,n)$ can be simply expressed in terms of the
propagator $P(x,A,n|x_0,x_0,0)$ computed above (\ref{start_expr}) as (see Fig. \ref{fig_rw} b)): 
\begin{eqnarray}\label{start_free}
\fl \tilde P(x,m | A,n) = \frac{1}{P(A,n)} \int_{-\infty}^\infty \rmd x_F \int_{-\infty}^\infty \rmd A_1 P(x,A_1,m|0,0,0) P(x_F,A-A_1,n-m|x,x,0) \;,
\end{eqnarray}
where we have used the Markov property of the L\'evy random walk. In the above expression (\ref{start_free}), $x_F$ is the end point of the walk (see Fig. \ref{fig_rw} b)), which is free here. Hence $\tilde P(x,m | A,n)$ is obtained by integration over this end point $x_F$. Notice that it is normalized according to 
$\int_{-\infty}^\infty   \tilde P(x,m | A,n) \rmd x = 1$ (and therefore we have divided by $P(A,n)$ in the expression above (\ref{start_free}) because the measure is restricted to random walks of fixed area $A$ after $n$ time steps). Using the explicit expressions computed above (\ref{start_expr}) one obtains after integration over $x_F$ and $A_1$:
\begin{eqnarray}
 \fl \tilde P(x,m | A,n) = \frac{1}{P(A,n|0,0)} \int_{-\infty}^\infty \frac{\rmd k_1}{2\pi} \int_{-\infty}^\infty \frac{\rmd k_2}{2\pi} e^{-\sum_{\nu=0}^m |k_1+\nu k_2|^\alpha - |k_2|^\alpha \sum_{\nu=0}^{n-m} |\nu|^\alpha} e^{- i k_1 x -ik_2 (A - (n-m) x)} \;. \nonumber \\
\end{eqnarray}
In the large $n$ limit, keeping $X = x/n^{1/\alpha}$, $Y = A/n^{1+1/\alpha}$ and $\tau = m/n$ fixed one has
\begin{eqnarray}
\fl && \tilde P(x,m | A,n) = \frac{1}{n^{1/\alpha}} \tilde G(X, \tau | Y) \;,  \nonumber \\
\fl && \tilde G(X, \tau | Y) = \frac{(1+\alpha)^{-1/\alpha}}{{\cal S}_\alpha((\alpha + 1)^{1/\alpha} Y)}  \int_{-\infty}^\infty \frac{\rmd r}{2\pi} \int_{-\infty}^\infty \frac{\rmd k}{2\pi}  |k| e^{-|k|^\alpha\left( \tilde \gamma(r,\tau) + \tilde \gamma(0,1-\tau) \right)} e^{- i k r X - i k (Y - (1-\tau) X)} \;,
\end{eqnarray}
where we have used the expression of $P(A,n|0,0)$ given in Eq. (\ref{marginals}) and we have introduced 
\begin{eqnarray}
\tilde  \gamma(r,\tau) = \int_0^\tau |r + z|^\alpha dz \;,
\end{eqnarray}
which is a generalization of the function $\gamma(r) \equiv \tilde \gamma(r,1)$ in (\ref{expr_scaling}). It reads
\begin{eqnarray}\label{expr_gen_gamma}
 \tilde \gamma(r , \tau) =
\cases{
\frac{1}{\alpha+1} \left( (-r)^{\alpha+1} - (-\tau-r)^{\alpha+1} \right) \;, \; r \leq -\tau \;, \\
\frac{1}{\alpha+1} \left( (r+\tau)^{\alpha+1} + (-r)^{\alpha+1}  \right) \;, \; -\tau \leq r \leq 0 \;, \\
\frac{1}{\alpha+1} \left( (r+\tau)^{\alpha+1} - r^{\alpha+1}  \right) \;, \; r \geq 0 \;.
}
\end{eqnarray}
We can now compute $\langle \tilde x(m) \rangle$ which, in the large $n$ limit, takes the scaling form 
\begin{eqnarray}\label{free_1}
\fl &&\frac{\langle \tilde x(m) \rangle}{n^{1/\alpha}} = h_\alpha\left(\frac{m}{n}, \frac{A}{n^{1+1/\alpha}}\right)  \\
\fl &&h_\alpha(\tau, Y) =  \frac{(1+\alpha)^{-1/\alpha}}{{\cal S}_\alpha((\alpha + 1)^{1/\alpha} Y)} \int_{-\infty}^\infty \rmd X X \int_{-\infty}^\infty \frac{\rmd r}{2\pi} \int_{-\infty}^\infty \frac{\rmd k}{2\pi}  |k| e^{-|k|^\alpha g(r,\tau)} e^{- i k r X - i k (Y - (1-\tau) X)} \;. \nonumber 
\end{eqnarray}
with $g(r,\tau) = \tilde \gamma(r,\tau) + \tilde \gamma(0,1-\tau)$. This function $h_\alpha(\tau, Y)$ can be written as
\begin{eqnarray}
 \fl &&h_\alpha(\tau, Y) =  \frac{(1+\alpha)^{-1/\alpha}}{{\cal S}_\alpha((\alpha + 1)^{1/\alpha} Y)} \int_{-\infty}^\infty \rmd X \int_{-\infty}^\infty \frac{\rmd r}{2\pi} \int_{-\infty}^\infty \frac{\rmd k}{2\pi} \frac{|k|}{(- i k)} e^{-|k|^\alpha g(r,\tau)} \frac{\partial}{\partial r} \left( e^{- i k r X - i k (Y - (1-\tau) X)}\right)
\end{eqnarray}
which suggests to perform an integration by part in the integral over $r$, yielding (one can check that the boundary terms vanish)
\begin{eqnarray}\label{intermediaire}
 \fl h_\alpha(\tau, Y) =  \frac{(1+\alpha)^{-1/\alpha}}{{\cal S}_\alpha((\alpha + 1)^{1/\alpha} Y)} \int_{-\infty}^\infty \rmd X \int_{-\infty}^\infty \frac{\rmd r}{2\pi} \int_{-\infty}^\infty \frac{\rmd k}{2\pi} \frac{|k|^{1+\alpha}}{(- i k)}\frac{\partial g(r,\tau)}{\partial r} e^{-|k|^\alpha g(r,\tau)}  e^{- i k r X - i k (Y - (1-\tau) X)} \nonumber \\
\end{eqnarray}
On this expression (\ref{intermediaire}), the integral over $X$ can be done yielding simply a delta function of $r$, namely $2 \pi |k|^{-1}\delta(r - (1-\tau))$. This allows us to perform then the integral over $r$ to obtain
\begin{eqnarray}\label{free_2}
&& h_\alpha(\tau, Y) =  \frac{(1+\alpha)^{-1/\alpha}}{{\cal S}_\alpha((\alpha + 1)^{1/\alpha} Y)} i \left(\frac{\partial g(r,\tau)}{\partial r} \right)_{r = 1-\tau}
\int_{-\infty}^\infty  |k|^\alpha k^{-1} e^{-\frac{|k|^\alpha}{1+\alpha}} e^{-ikY} \frac{\rmd k}{2\pi} \;,
\end{eqnarray}
where we have used the relation $\tilde \gamma(1-\tau,\tau) + \tilde \gamma(0,1-\tau) = 1/(1+\alpha)$. It is then easy to check that
\begin{eqnarray}
 i \int_{-\infty}^\infty |k|^\alpha k^{-1} e^{-\frac{|k|^\alpha}{1+\alpha}} e^{-ikY} \frac{\rmd k}{2\pi} = \frac{\alpha+1}{\alpha} Y (1+\alpha)^{1/\alpha} {\cal S}_\alpha\left[(1+\alpha)^{1/\alpha} Y\right]  \;,
\end{eqnarray}
so that finally one obtains the simple result
\begin{eqnarray}\label{explicit_free}
 h_\alpha(\tau, Y) = Y \frac{\alpha+1}{\alpha} (1- (1-\tau)^\alpha) \;,
\end{eqnarray}
where we have used $\left(\frac{\partial g(r,\tau)}{\partial r} \right)_{r = 1-\tau} = 1-(1-\tau)^\alpha$. In Fig. (\ref{fig_profiles}) a), we show a plot of $h_\alpha(\tau, Y)/Y$ as a function of $\tau$.  

\subsection{The case of a L\'evy bridge}

Here we consider a L\'evy bridge, i.e. a L\'evy random walker starting in $0$ at initial time and constrained to come back to the origin
after $n$ time steps. We compute the probability $\tilde P_B(x,m |
A,n)$ that the position of the random walker is $x$ after $m$ time
steps given that the area, after $n$ time steps, is fixed to $A$. From
this probability, one obtains the average profile as $\langle \tilde x_B(m)
\rangle = \int_{-\infty}^\infty x \tilde P_B(x,m | A,n) \rmd
x$. This probability $\tilde P_B(x,m | A,n)$ can be expressed, as in Eq. (\ref{start_free}) in terms
of the propagator $P(x,A,n| x_0,x_0,0)$ computed above in Eq. (\ref{start_expr}) as:
\begin{eqnarray}\label{start_dist_profile}
\fl \tilde P_B(x,m | A,n) = \frac{1}{P(0,A,n|0,0,0)}\int_{-\infty}^\infty  P(x,A_1,m|0,0,0) P(x,A-A_1,n-m|0,0,0) \rmd A_1 \;,
\end{eqnarray}
where we have used the Markov property of the L\'evy random walk. It
is normalized according to $\int_{-\infty}^\infty \tilde P_B(x,m | A,n) \rmd
x = 1$ (and therefore we have divided by $P(0,A,n|0,0,0)$ because the
measure is restricted to bridges of fixed area $A$). Using the explicit
expressions obtained above (\ref{start_expr}) one has 
\begin{eqnarray}\label{dist_profile}
\tilde P_B(x,m | A,n)  =  &&\frac{1}{P(0,A,n|0,0,0)}\int_{-\infty}^\infty \frac{\rmd k_1}{2\pi} \int_{-\infty}^\infty \frac{\rmd k'_1}{2\pi} \int_{-\infty}^\infty \frac{\rmd k_2}{2\pi} e^{-i(k_1 + k'_1) x - i k A} \nonumber \\
&& \times
e^{- \sum_{\nu=1}^n |k_1 + \nu k_2|^\alpha - \sum_{\nu=1}^{n-m} |k'_1 + \nu k_2|^\alpha} \;.
\end{eqnarray}
In the large $n$ limit, keeping $X = x/n^{1/\alpha}$, $Y = A/n^{1+1/\alpha}$ and $\tau = m/n$ fixed one has
\begin{eqnarray}\label{start_expr_profile}
&& \tilde P_B(x,m | A,n) =  \frac{1}{n^{1/\alpha}} \tilde G_B(X,\tau|Y) \;, \nonumber \\
&& \tilde G_B(X,\tau|A) = \frac{\pi}{\Gamma(1+\alpha^{-1}) F_\alpha(Y)}\int_{-\infty}^\infty \frac{\rmd r}{2\pi} \int_{-\infty}^\infty \frac{\rmd r'}{2\pi}\int_{-\infty}^\infty \frac{\rmd k}{2\pi} k^2 e^{-i k (r + r') X - i k A} \\
&& \times e^{ - |k|^\alpha [\tilde \gamma(r,\tau) + \tilde \gamma(r',1-\tau) ] }
\end{eqnarray}

We can now compute $\langle \tilde x_B(m) \rangle$, which in the large $n$ limit takes the scaling form
\begin{eqnarray}
\frac{\langle \tilde x_B(m) \rangle}{n^{1/\alpha}} = H_\alpha\left(\frac{m}{n}, \frac{A}{n^{1+1/\alpha}} \right) \;,
\end{eqnarray}
where
\begin{eqnarray}\label{expr_bridge_stable}
\!\!\!\!\!H_\alpha(\tau, Y) &=& i \frac{1}{2\Gamma(1+\alpha) F_\alpha(Y)} \int_{-\infty}^\infty \frac{\rmd k}{2 \pi} \int_{-\infty}^\infty {\rmd} r |k|^{\alpha-1} k \partial_r \tilde \gamma(r,\tau) e^{- |k|^\alpha \tilde \Gamma(r,\tau) - i k Y} \\
\!\!\!\!\!\! &=& - \frac{1}{2\Gamma(1+\alpha) F_\alpha(Y)}  \frac{\partial}{\partial Y} \left( \int_{-\infty}^\infty \frac{\rmd k}{2 \pi} \int_{-\infty}^\infty {\rmd} r |k|^{\alpha-1}
\partial_r \tilde \gamma(r,\tau) e^{- |k|^\alpha \tilde \Gamma(r,\tau) - i k Y} \right)\;, \nonumber
\end{eqnarray}
where we have introduced the notation $\tilde \Gamma(r,\tau) = \tilde \gamma(r,\tau) + \tilde \gamma(-r,1-\tau)$ which we compute
straightforwardly from Eq. (\ref{expr_gen_gamma}) as:
\begin{eqnarray}\label{def_biggamma}
 \fl \tilde \Gamma(r,\tau) = \tilde \gamma(r,\tau) + \tilde \gamma(-r,1-\tau) = 
\cases{
\frac{1}{\alpha+1} \left((-r+1-\tau)^{\alpha+1} - (-r-\tau)^{\alpha+1} \right) \;, \; r \leq -\tau \\
\frac{1}{\alpha+1} \left( (r+\tau)^{\alpha+1} + (-r+1-\tau)^{\alpha+1}   \right) \;, \; -\tau \leq r \leq 1-\tau \\
\frac{1}{\alpha+1} \left( (r+\tau)^{\alpha+1} - (r-1+\tau)^{\alpha+1}  \right) \;, \; r \geq 1-\tau \;.
}
\end{eqnarray}
Note that this function $\tilde \Gamma(r,\tau)$ satisfies the identity
\begin{eqnarray}
\tilde \Gamma(1-\tau-r,\tau) = \int_0^{1} |z-r|^\alpha \rmd z = \gamma(-r) \;,
\end{eqnarray}
independently of $\tau$. 

For generic $\alpha$, the expression above (\ref{expr_bridge_stable})
is quite difficult to handle. For $\alpha = 2$ (Brownian motion) and
$\alpha = 1$, further analytical progress is however possible. For
$\alpha = 2$, one has $\partial_r \tilde \gamma(r,\tau) =
\tau(2r+\tau)$ and $\tilde \Gamma(r,\tau) = r^2 + r (2\tau-1) -
\tau(1-\tau)+1/3$ and therefore one checks
\begin{eqnarray}\label{identity}
 \int_{-\infty}^\infty  \partial_r \tilde \gamma(r,\tau) e^{-k^2
 \tilde \Gamma(r,\tau) } \rmd r = \frac{\sqrt{\pi}}{|k|} e^{-\frac{k^2}{12}} \tau(1-\tau) \;.
\end{eqnarray}
Using this identity (\ref{identity}) and integrating over $r$ in the expression above (\ref{expr_bridge_stable}), and using $F_2(Y) = \sqrt{3/\pi}\exp{(-3Y^3)}$ one obtains for $\alpha=2$:
\begin{eqnarray}
 H_2(\tau, Y) = 6 Y \tau(1-\tau) \;.
\end{eqnarray}

Another interesting case where analytical progress is possible is $\alpha =1$. Given the expression of $\tilde \gamma(r,\tau)$ in Eq. (\ref{expr_gen_gamma}) and $\tilde \Gamma(r,\tau)$ in Eq. (\ref{def_biggamma}), one 
observes that the integral over $r$ in Eq. (\ref{expr_bridge_stable}) gives rise to $4$ different terms, corresponding to $r \in ]-\infty, -\tau]$, $r \in [-\tau,0]$, 
$r \in [0, 1-\tau]$ and finally $r \in [1-\tau,+\infty[$. For $\alpha = 1$ it turns out that the first and fourth terms, corresponding to  $r \in ]-\infty, -\tau]$ and $r \in [1-\tau, +\infty[$ do cancel each other (which is the case only for $\alpha=1$) resulting in the following expression:
\begin{eqnarray}
\fl H_1(\tau, Y) = \frac{Y}{\pi \Gamma(1+\alpha) F_1(Y)} \left(
\int_{-\tau}^0  (2r+\tau) \frac{\tilde \Gamma(r,\tau)}{(\tilde \Gamma(r,\tau)^2 +
  Y^2)^2} \rmd r + \tau \int_{0}^{1-\tau}
\frac{\tilde \Gamma(r,\tau)}{(\tilde \Gamma(r,\tau)^2+Y^2)^2} \rmd r
\right) \;,
\end{eqnarray}
with $\tilde \Gamma(r,\tau) = \frac{1}{2} ((r+\tau)^2+(r+\tau-1)^2)$. In the asymptotic limit $Y \to 0$ one obtains
\begin{eqnarray}\label{small_y_shape}
 H_1(\tau, Y) \sim \frac{Y}{4 + \pi} \left( 3\pi-2 + \frac{2}{1+2\tau(\tau-1)} + 12 (2\tau-1) \arctan{(1-2\tau)} \right) \;.
\end{eqnarray}
In the opposite limit $Y \to \infty$ one obtains
\begin{eqnarray}\label{large_y_shape}
 H_1(\tau, Y) \sim  Y \frac{10}{3} \tau(1-\tau)(2 - \tau(1-\tau)) \;.
\end{eqnarray}
Note that although the two functions of $\tau$ entering these asymptotic expansions in Eq. (\ref{small_y_shape}) and Eq. (\ref{large_y_shape}) have very different analytical expressions, they are actually quite close to each other on the interval $[0,1]$ (see Fig. (\ref{fig_profiles}) b)).

\section{Numerical results}

\begin{figure}[ht]
\begin{center}
 \includegraphics [width = \linewidth]{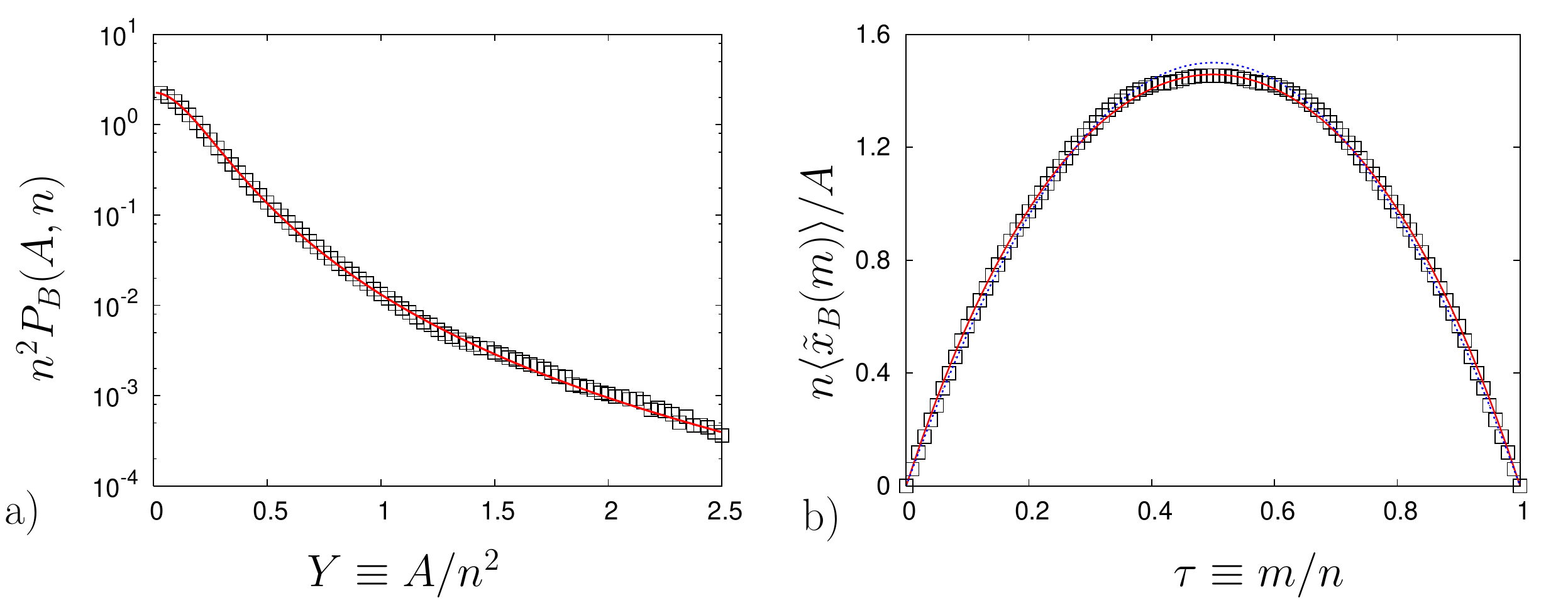}
 \end{center}
\caption{{\bf a)}: The squares represent the numerical data for $n^2 P_B(A,n)$ as a function of $Y \equiv A/n^2$ for a L\'evy bridge of index $\alpha=1$ of length $n=100$. The solid line is our exact expression for $F_1(Y)$ given in Eq. (\ref{elem}). {\bf b)}: The squares represent our numerical data for $n \langle \tilde x_B(m) \rangle/A$ as a function of $\tau =m/n$ for a L\'evy bridge of index $\alpha=1$ of length $n=100$ and $A/n^2 = 20$. The solid line corresponds to our asymptotic result
for large $A$ given in Eq. (\ref{large_y_shape}) while the dotted line represents the result for the Brownian bridge in Eq. (\ref{shape_bb}).}\label{fig_numerics}
\end{figure}

We now come to numerical simulations of L\'evy bridges. As mentioned above, one can not use the relation above (\ref{id_bb}), which is only valid for $\alpha=2$ \cite{knight} to simulate a L\'evy bridge. Instead, we consider the joint probability distribution function (pdf) of the increments $\eta(m)$ for a L\'evy bridge of size $n$. Indeed, these increments are independent random variables, distributed
according to $\phi(\eta)$ with the global constraint that $x(n) = \sum_{m=1}^n \eta(n) = 0$. Therefore the joint pdf of the increments $P_B\left(\eta(1), \eta(2), \cdots, \eta(n)\right)$ is simply given by
\begin{eqnarray}\label{joint_increments}
P_{B}\left(\eta(1), \eta(2), \cdots, \eta(n) \right) &\propto& \prod_{m=1}^n \phi\left[\eta(m)\right] \delta\left(\sum_{m=1}^n \eta(m)\right) \\
&\propto& \exp{\left[ \sum_{m=1}^n \ln{\left [\phi\left[\eta(m)\right] \right]} \right ]} \delta \left(\sum_{m=1}^n \eta(m) \right) \;. \nonumber 
\end{eqnarray}
This joint distribution can thus be considered as a Boltzmann weight with an effective
energy $E = -  \sum_{m=1}^n \ln{\left [\phi\left[\eta(m)\right] \right] }$ and effective inverse temperature $\beta = 1$. This thus leads us to use a Monte-Carlo algorithm, with a global constraint, to generate "configurations" of the increments distributed according to the distribution above (\ref{joint_increments}). We implement it in the following way. We start with a random initial configuration of the $\eta(m)$'s which satisfies the global constraint $\sum_{m=1}^n \eta(m)=0$ (it can also be $\eta(m) = 0$, for all $m$). At each time step we choose randomly two sites $i$ and $j$ among $1, 2, \cdots, n$ and the simple following moves are proposed 
\begin{eqnarray}\label{move}
&&\eta(i) \to \eta'(i) = \eta(i)  + \Delta \eta \;, \nonumber \\
&&\eta(j) \to \eta'(j) = \eta(j)  - \Delta \eta  \;,
\end{eqnarray} 
such that the global constraint of zero sum is automatically satisfied. This move is then accepted, in Metropolis
algorithm that we use here, with a probability $P_{ij}$ given by
\begin{eqnarray}
P_{ij} &=& \min{\left(1, \frac{\phi \left[ \eta'(i) \right] \phi\left[\eta'(j)\right]}{\phi \left[\eta(i) \right] \phi\left[\eta(j) \right]} \right)} \\
&=& \min(1,\exp{(-\Delta E)}) \;, \; \Delta E = \log{\left(\frac{\phi \left[\eta(i) \right] \phi\left[\eta(j) \right]}{\phi \left[ \eta'(i) \right] \phi\left[\eta'(j)\right]}\right)}
\end{eqnarray}
This Monte Carlo algorithm is thus very similar to the Kawasaki dynamics for ferromagnetic spin systems relaxing
towards equilibrium with a conserved global magnetization \cite{kawasaki}. Once the increments $\eta(k)$'s are generated according to this joint
probability (\ref{joint_increments}), we can generate the random walk bridge $x_B(m) = \sum_{k=1}^m \eta(k)$ and compute the distribution of the area $A = \sum_{m=1}^n x_B(m)$ under the L\'evy bridge. In Fig. \ref{fig_numerics} a), we show a plot of this distribution $P_B(A,n)$ for $\alpha=1$ and $n=100$. To compute it we have first run $10^7$ Monte Carlo steps to equilibrate the system and the distribution was then computed as an average over $10^7$ samples generated in the time interval $[10^7, 2 . 10^7]$. In Fig. \ref{fig_numerics}, we also show a plot of the exact explicit expression for $F_1(Y)$ given in Eq. (\ref{elem}), showing a very good agreement with our numerics. We have also computed numerically this distribution for other values of $\alpha \in ]0,2[$, showing a good agreement with the power law tail obtained in Eq. (\ref{large_y}). Note however that for small $\alpha$, it is actually quite difficult to equilibrate the system such that a precise estimate of the exponent characterizing the power law tail of $P_B(A,n)$ is quite difficult for $\alpha < 1$.  

We can use a similar Monte Carlo approach to generate a random walk bridge with a fixed area $A = \sum_{m=1}^n (n+1-m) \eta(m) $. In that case, the joint
pdf of the increments $\tilde P_B\left(\eta(1), \eta(2), \cdots, \eta(n)\right)$ is simply given by
\begin{eqnarray}\label{joint_increments_area}
\fl P_{B}\left(\eta(1), \eta(2), \cdots, \eta(n) \right) &\propto& \prod_{m=1}^n \phi\left[\eta(m) \right ]\delta\left(\sum_{m=1}^n \eta(m)\right) \delta(\sum_{m=1}^n (n+1-m) \eta(m)-A) \;.
\end{eqnarray}
We start with an initial configuration of the $\eta(m)$'s which satisfies the both global constraints. In practice, we start with $\eta(m) = 6 A (N+1-2 m)/((N-1)N(N+1))$. Then, to satisfy both constraints (\ref{joint_increments_area}), at each time step we choose randomly three sites $i$, $j$ and $k$ among $1, 2, \cdots, n$ and the simple following moves are proposed 
\begin{eqnarray}\label{move_area}
&&\eta(i) \to \eta'(i) = \eta(i) + \Delta \eta \;, \nonumber \\
&&\eta(j) \to \eta'(j) = \eta(j) + \frac{i-k}{k-j} \Delta \eta \;, \nonumber \\
&&\eta(k) \to \eta'(k) = \eta(k) + \frac{j-i}{k-j} \Delta \eta \;. 
\end{eqnarray}
Note that to converge to the correct probability measure (\ref{joint_increments_area}) one has to choose $\eta$ either positive or negative with equal probability. 
This move (\ref{move_area}) is then accepted with a probability $P_{ijk}$ given by
\begin{eqnarray}
P_{ijk} &=& \min{\left(1, \frac{\phi\left[\eta'(i)\right] \phi\left[\eta'(j)\right] \phi\left[\eta'(k)\right]}{\phi\left[ \eta(i)\right] \phi\left[\eta(j)\right] \phi\left[\eta(k) \right]} \right)} \\
&=& \min(1,\exp{(-\Delta E)}) \;, \; \Delta E = \log{\left(\frac{\phi \left[\eta(i) \right] \phi\left[\eta(j) \right] \phi\left[\eta(k) \right]  }{\phi \left[ \eta'(i) \right] \phi\left[\eta'(j)\right] \phi\left[\eta'(k) \right] }\right)}
\end{eqnarray}
Once the increments $\eta(k)$'s are generated according to this joint
probability (\ref{joint_increments_area}), we can generate the random walk bridge $\tilde x_B(m) = \sum_{k=1}^m \eta(k)$ with fixed area $A$ and compute
the profile $\langle \tilde x_B(m) \rangle$. In Fig.~\ref{fig_numerics} b), we show a plot of this average profile for $\alpha=1$, $n=100$ and $A/n^2 \sim 20$. To compute it we have first run $10^7$ Monte Carlo steps to equilibrate the system and the average was then computed over $10^7$ samples generated in the time interval $[10^7, 2 . 10^7]$. In Fig. \ref{fig_numerics} b), we also plot, with a solid line, our asymptotic result in Eq. (\ref{large_y_shape}), showing a relatively good agreement with our numerics (note that here $A/n^2 = 20$). On the same plot, Fig. \ref{fig_numerics} b), we also show in dotted line, the result for the Brownian bridge (\ref{shape_bb}), which is independent of $A$. It is quite remarkable that these two profiles are very similar which show that the global constraints that we impose here have strong consequences on the statistics of the L\'evy random walk.

\section{Conclusion}

To conclude, we have studied two main properties of a L\'evy bridge $x_B(m)$ of length $n$ : (i) the distribution $P_B(A,n)$ of the area under a L\'evy bridge and (ii) the average profile $\langle \tilde x_B(m) \rangle$ of a L\'evy bridge with fixed area $A$. 
\begin{itemize}
\item{
For $P_B(A,n)$ we have found the scaling form, valid for large $n$, $P_B(A,n) \sim n^{-1-1/\alpha} F_\alpha(Y)$ with an interesting power law behavior $F_\alpha(Y) \sim Y^{-2(1+\alpha)}$. For $\alpha=1$, we have obtained an explicit expression for $F_1(Y)$ in terms of elementary functions (\ref{elem}). We have also shown, using the Lindeberg condition that the non-Gaussianity of $P_B(A,n)$, for $\alpha > 1/2$, is due only to the correlations between the positions of the walkers $x_B(m)$'s.}
\item{For the average profile, $\langle \tilde x_B(m) \rangle$, we have found the scaling form $\langle \tilde x_B(m) \rangle \sim n^{1/\alpha} H_{\alpha}(m/n,A/n^{1+1/\alpha})$ where, at variance with Brownian motion, $H_\alpha(X,Y)$ is a non trivial function of the rescaled area $Y$. For $\alpha=1$, we have obtained simple analytical expressions for $H_1(X,Y)$ in both limits $Y \to 0$ and $Y \to \infty$. In particular, we have shown that the average profile of the L\'evy random walk with a fixed area is not very far from the profile of a Brownian bridge with fixed area.}
\item{We have finally compared our analytical results with Monte Carlo simulations of these L\'evy random walks with global constraints.}
\end{itemize}
In view of recent developments in the study of area distributions for variants of Brownian motions \cite{janson_review, satya_airy, schehr_airy, rambeau_airy}, it would be very interesting to extend the results presented here to other constrained L\'evy walk, including in particular L\'evy random walks conditioned to stay positive (L\'evy excursion), which is a challenging open problem.

\newpage

\appendix

\section{On the use of the Lindeberg condition}\label{appendix_lindeberg}

Let us consider {\it independent} and {\it non-identical} random variables $X_1, \cdots, X_n$ which have the same distribution as the L\'evy bridge $x_B(m)$ (\ref{marginal_bridge1}), {\it i.e.}
\begin{eqnarray}
\!\!\!\! {\rm Proba}(X_m = x) = \frac{\pi}{\Gamma(1+\alpha^{-1/2})} \frac{n^{1/\alpha}}{m^{1/\alpha}(n-m)^{1/\alpha}} {\cal S}_\alpha \left(\frac{x}{m^{1/\alpha}} \right) {\cal S}_\alpha \left(\frac{x}{(n-m)^{1/\alpha}} \right) \;.
\end{eqnarray}
For $\alpha > 1/2$, $\sigma_m^2 = \langle X_m^2 \rangle$ is well defined and one has (\ref{variance_bridge})
\begin{eqnarray}
 \sigma_m^2 = \langle X_m^2 \rangle = \langle x^2_B(m) \rangle = \tilde a_\alpha n^{2/\alpha - 2} m (n-m)  \;.
\end{eqnarray}
Given that the variables $X_i$ are not identical, one can not apply directly the Central Limit Theorem. However, one can show that these
random variables $X_m$ do satisfy the Lindeberg condition which guarantees that their sum $A_n = \sum_{m=0}^n X_m$ is distributed according
a Gaussian distribution in the large $n$ limit. Let us first introduce $\Sigma_n^2$
\begin{eqnarray}
 \Sigma_n^2 = \sum_{m=1}^n \sigma_m^2 = \frac{\tilde a_\alpha}{2} n^{2/\alpha-2} (n+1) n (n-3) \sim \frac{\tilde a_\alpha}{2} n^{2/\alpha +1} \;, \; n \gg 1 \;,
\end{eqnarray}
which implies $\Sigma_n \sim n^{1/2+1/\alpha}$ for large $n$. To apply the Lindeberg condition, we need to estimate for any $\epsilon > 0$
\begin{eqnarray}
&& \langle X_m^2 \rangle_\epsilon = \int_{|x| > \epsilon \Sigma_n} x^2 {\rm Proba}(X_m = x) \rmd x \\
&& = 2  \frac{\pi}{\Gamma(1+\alpha^{-1/2})} \frac{n^{1/\alpha}}{m^{1/\alpha}(n-m)^{1/\alpha}} \int_{\epsilon \Sigma_n}^\infty  x^2 {\cal S}_\alpha \left(\frac{x}{m^{1/\alpha}} \right) {\cal S}_\alpha \left(\frac{x}{(n-m)^{1/\alpha}} \right)  \rmd x \nonumber \\
&& \sim c'_\alpha n^{2/\alpha} n^{-(3/2+\alpha)} m (n-m) \;,
\end{eqnarray}
where $c'_\alpha$ is independent of $m$ and $n$. Therefore one has
\begin{eqnarray}\label{last_lindeberg}
 \frac{\sum_{m=0}^n \langle X_m^2 \rangle_\epsilon}{\Sigma_n^2} \sim n^{-(\alpha-1/2)} \;.
\end{eqnarray}
For random variables for which the above ratio (\ref{last_lindeberg}) goes to zero in the limit $n \to \infty$ (which is the case here for $\alpha < 1/2$), 
a theorem due to Lindeberg (thus called the 'Lindeberg condition') \cite{feller}, says that their sum $A_n/\Sigma_n = \Sigma_n^{-1}\sum_{m=0}^n X_m$ is distributed, in the limit $n \to \infty$, according to a Gaussian distribution of unit variance. The fact that, for a L\'evy bridge, the area is not a Gaussian distribution (\ref{large_y}) is thus, for $\alpha > 1/2$, a consequence of the correlations between the random variables $x_B(m)$. 

To conclude this paragraph, we discuss a simple case where the Lindeberg condition does not hold. Consider the case where $X_1, \cdots, X_n$ are independent random variables distributed according to~\cite{bertin_review}
\begin{eqnarray}
 {\rm Proba}(X_m=x) = m e^{-mx} \;,
\end{eqnarray}
such that one has
\begin{eqnarray}
 \langle X_m \rangle = \frac{1}{m} \;, \; \langle (X_m - \langle X_m \rangle)^2 \rangle = \frac{1}{m^2} \;.
\end{eqnarray}
Then in that case one has immediately
\begin{eqnarray}
 \Sigma_n^2 = \sum_{m=1}^n  \langle (X_m - \langle X_m \rangle)^2 \rangle  = \sum_{m=1}^n \frac{1}{m^2} \to \frac{\pi^2}{6} \;, \; n \to \infty \;. 
\end{eqnarray}
One computes straightforwardly, for $\epsilon > 0$
\begin{eqnarray}
 \int_{\epsilon \Sigma_n}^\infty (X_m - \langle X_m \rangle)^2 {\rm Proba}(X_m=x) \rmd x = \frac{e^{-\epsilon \Sigma_n m}}{m^2} \left( 1 + m^2 (\epsilon \Sigma_n)^2 \right) \;,
\end{eqnarray}
such that here one has
\begin{eqnarray}
\hspace*{-1cm} \frac{1}{\Sigma_n^2} \sum_{m=1}^\infty \int_{\epsilon \Sigma_n}^\infty (X_m - \langle X_m \rangle)^2 {\rm Proba}(X_m=x) \rmd x \to \frac{6}{\pi^2} \sum_{m=1}^\infty \frac{e^{-\epsilon' m}}{m^2} (1+ (\epsilon' m)^2) > 0 \;, 
\end{eqnarray}
with $\epsilon' = \epsilon \pi^2/6$. Therefore the Lindeberg condition (\ref{lindeberg}) does not hold here. In fact, it can be shown that the distribution
of the variable $(S_n- \sum_{k=1}^n k^{-1})/\Sigma_n$ converges to a Gumbel distribution \cite{bertin_review}.

\section{Asymptotic behavior of $F_\alpha(Y)$ for large $Y$}\label{app_asympt}

To analyse the large argument behavior of $F_\alpha(Y)$, we analyse the small $k$ behavior of its Fourier transform $\hat F_\alpha(k)$ given in the text in Eq. (\ref{expr_fourier_gen_alpha}): 
\begin{eqnarray}
\hat F_\alpha(k) &=& \int_{-\infty}^\infty \rmd Y F_\alpha(Y) e^{i k Y}  = \hat F_{\alpha,1}(k) +\hat F_{\alpha,2}(k) \;, \\
\hat F_{\alpha,1}(k) &=& \frac{|k|}{\Gamma(1+\alpha^{-1})} \int_0^\infty e^{-\frac{|k|^\alpha}{\alpha+1} \left[(r+1)^{\alpha+1}-r^{\alpha+1}\right]}  \rmd r \;, \\
\hat F_{\alpha,2}(k) &=& \frac{|k|}{\Gamma(1+\alpha^{-1})} \int_0^{1/2}    e^{-\frac{|k|^\alpha}{\alpha+1} \left[(1/2+r)^{\alpha+1}+(1/2-r)^{\alpha+1}\right]} \rmd r \;. 
\end{eqnarray}
The analysis of the small $k$ behavior of $\hat F_{\alpha,2}(k)$ is simply obtained by expanding the exponential under the integral. It yields straightforwardly:
\begin{eqnarray}\label{f2_small_k}
\fl \hat F_{\alpha,2}(k) = \frac{1}{\Gamma(1+\alpha^{-1})} \left[\frac{|k|}{2} - \frac{|k|^{1+\alpha}}{(\alpha+1)(\alpha+2)} + \frac{|k|^{1+2\alpha}}{2 (\alpha+1)^2} \left( \frac{1}{3+2 \alpha} + \frac{\sqrt{2 \pi} \Gamma(2+\alpha)}{\Gamma(\frac{5}{2}+\alpha) 2^{3 + 2\alpha}} \right)
\right] + {\cal O}(|k|^{1+3\alpha}) \nonumber \\
\end{eqnarray}

The asymptotic expansion of $\hat F_{\alpha,1}(k)$ is a bit a more subtle. To get the two first terms of the expansion, one performs the change of variable $z = |k| r$ and then expand $(r+1)^{\alpha+1}$ using the binomial formula, 
\begin{eqnarray}
 \frac{1}{\alpha+1} ((r+1)^{\alpha+1}-r^{\alpha+1} ) = r^{\alpha} + \frac{\alpha}{2} r^{\alpha-1} + ... \;.
\end{eqnarray}
This yields
\begin{eqnarray}\label{intermediate_f1}
 \hat F_{\alpha,1}(k)  \sim \frac{1}{\Gamma(1+\alpha^{-1})} \int_0^{\infty} e^{-r^{\alpha} - \frac{\alpha}{2} |k| r^{\alpha-1}} \rmd r + {\cal O}(k^{1+\eta}) \;,
\end{eqnarray}
where $\eta > 0$ is yet unknown (see below). From this expression (\ref{intermediate_f1}), one immediately obtains the two first terms of the expansion of $\hat F_{\alpha,1}(k)$ as
\begin{eqnarray}\label{f1_small_k_first}
 \hat F_{\alpha,1}(k) = 1 - \frac{|k|}{2\Gamma(1+\alpha^{-1})} + {\cal O}(k^{1+\eta}) \;.
\end{eqnarray}
 Combining Eq. (\ref{f2_small_k}) and Eq. (\ref{f1_small_k_first}) one sees that the first non-trivial term, proportional to $|k|$ cancel in $\hat F_\alpha(k)$. Therefore, one needs to develop $\hat F_{\alpha,1}(k)$ beyond the first terms (\ref{f1_small_k_first}). To this purpose, we need to study separately the cases $0 < \alpha \leq 1/2$, $1/2 < \alpha \leq 1$ and $1 < \alpha \leq 2$.

\subsection{The case $0 < \alpha \leq 1/2$}

Let us first analyse the term $\hat F_{\alpha,1}(k)$ which we decompose as
\begin{eqnarray}
&& \hat F_{\alpha,1}(k) = B_1 (k) + B_2 (k) + B_3(k) \label{decompose_app} \;, \\
&& B_1(k) = \frac{|k|}{\Gamma(1+\alpha^{-1})} \int_0^1 \rmd r e^{-\frac{|k|^\alpha}{\alpha+1} \left[(r+1)^{\alpha+1}-r^{\alpha+1}\right]}  \label{def_b1}\;, \nonumber \\
&& B_2(k) = \frac{|k|}{\Gamma(1+\alpha^{-1})} \left[ \int_1^\infty \rmd r e^{-\frac{|k|^\alpha}{\alpha+1} \left[(r+1)^{\alpha+1}-r^{\alpha+1}\right]}
- e^{-|k|^\alpha[r^{\alpha} + \frac{\alpha}{2}  r^{\alpha-1}]} \right] \label{def_b2} \;,  \\
&& B_3(k) = \frac{k}{\Gamma(1+\alpha^{-1})}\int_1^\infty \rmd r e^{-|k|^\alpha (r^{\alpha} + \frac{\alpha}{2} |k| r^{\alpha-1})} \label{def_b3} \;.
\end{eqnarray}
It is easy to expand $B_1(k)$ for small $k$ as
\begin{eqnarray}
\fl B_1(k) =\frac{|k|^{1+\alpha}}{\Gamma(1+\alpha^{-1})} \frac{2-2^{\alpha+2}}{(\alpha+1)(\alpha+2)} + \frac{|k|^{1+2\alpha}}{2 \Gamma(1+\alpha^{-1})} \int_0^1 \left[\frac{\rmd r}{\alpha+1} \left( (r+1)^{\alpha+1} - r^{\alpha+1}\right) \right]^2 + {\cal O}(k^{1+3 \alpha}) \nonumber \\
\end{eqnarray}
To expand $B_2(k)$, one checks the asymptotic behaviors, for large $r$
\begin{eqnarray}
&& \frac{1}{\alpha+1} [ (r+1)^{\alpha+1}-r^{\alpha+1} ] - (r^\alpha + \frac{\alpha}{2} r^{\alpha-1}) = {\cal O} (r^{\alpha - 2}) \;, \\
&& \left( \frac{1}{\alpha+1} [ (r+1)^{\alpha+1}-r^{\alpha+1} ]\right)^2 - (r^\alpha + \frac{\alpha}{2} r^{\alpha-1})^2 = {\cal O} (r^{2\alpha - 2}) \;,
\end{eqnarray}
 so that, for $\alpha < 1/2$ one can safely expand the exponentials in the integrand of $B_2(k)$ (\ref{def_b2}) up to second order to obtain
\begin{eqnarray}\label{exp_b2_1}
\fl B_2(k) = &-& \frac{|k|^{1+\alpha}}{\Gamma(1+\alpha^{-1})} \int_1^\infty \rmd r \left[ \frac{1}{\alpha+1} [ (r+1)^{\alpha+1}-r^{\alpha+1} ] - (r^\alpha + \frac{\alpha}{2} r^{\alpha-1}) \right] \\
\fl &+& \frac{|k|^{1+2\alpha}}{2 \Gamma(1+\alpha^{-1})} \int_1^\infty \rmd r \left( \left( \frac{1}{\alpha+1} [ (r+1)^{\alpha+1}-r^{\alpha+1} ]\right)^2 - (r^\alpha + \frac{\alpha}{2} r^{\alpha-1})^2 \right) \\
\fl &&+ {\cal O}(|k|^{\min(2, 1+3\alpha)}) \;.
\end{eqnarray}
Summing up the contributions from $B_1(k)$ and $B_2(k)$ and performing the integrals yields
\begin{eqnarray}\label{comb_b1_b2}
\fl B_1(k) + B_2(k) = &&\frac{|k|^{1+\alpha}}{\Gamma(1+\alpha^{-1})} \left(\frac{1}{(\alpha+1)(\alpha+2)} - \frac{1}{\alpha+1} - \frac{1}{2}\right)  \\
\fl && +\frac{|k|^{1+2\alpha}}{2\Gamma(1+\alpha^{-1})} \left[ \frac{1}{2} + \frac{1}{1+2\alpha} - \frac{1}{(1+\alpha^2) (3 + 2\alpha)} + \frac{\alpha^2}{8\alpha-4} + 2\frac{\Gamma(-3-2\alpha) \Gamma(1+\alpha)}{\Gamma(-\alpha)} \right] \nonumber \\
\fl && + {\cal O}(|k|^{\min(2, 1+3\alpha)}) \;,
\end{eqnarray}
which we have carefully checked using Mathematica. 

Let us now expand $B_3(k)$ (\ref{def_b3}) for small $k$. It is easily seen from Eq. (\ref{f1_small_k_first}) that the first terms of this expansion are indeed given by
\begin{eqnarray}\label{eq_b3_1}
 B_3(k) = 1 - \frac{|k|}{2\Gamma(1+\alpha^{-1})} + {\cal O}(|k|^{1+\mu}) \;,
\end{eqnarray}
with $\mu > 0$. To go beyond the lowest orders, we first perform a change of variable $x = |k| r$ and then compute $B_3''(k)$ and finally expand it for small $k$. This yields
\begin{eqnarray}\label{b3_seconde}
&& B_3''(k) = \frac{1}{\Gamma(1+\alpha^{-1})} \left(\alpha (\frac{1}{2} + (1 + \frac{\alpha}{2})) k^{\alpha-1} e^{-(1+\frac{\alpha}{2}) k^{\alpha}} + I_3(k)\right) \;, \\
&& I_3(k) = \left(\frac{\alpha}{2} \right)^2 \int_k^\infty  e^{-x^\alpha - \frac{\alpha}{2} k x^{\alpha-1}} x^{2(\alpha-1)}\, \rmd x \;,
\end{eqnarray}
where $2(\alpha - 1)<-1$ for $\alpha < 1/2$. One then obtains the small $k$ behavior of $I_3(k)$ by simply expanding the term $e^{- \frac{\alpha}{2} k x^{\alpha-1}}$ in the integrand. This yields, to lowest order
\begin{eqnarray}\label{eq_i3}
 I_3(k) = \frac{1}{1-2\alpha} \left(\frac{\alpha}{2} \right)^2 k^{2\alpha-1} + {\cal O}(k^{3\alpha-1}) \;.
\end{eqnarray}
From Eq. (\ref{b3_seconde}) and Eq. (\ref{eq_i3}), one obtains straightforwardly
\begin{eqnarray}\label{eq_b3_asympt}
&& B_3(k) = 1 - \frac{|k|}{\Gamma(1+\alpha^{-1})} + \frac{|k|^{1+\alpha}}{\Gamma(1+\alpha^{-1})} \left(\frac{1}{2} + \frac{1}{\alpha+1} \right) \\
&& + \frac{|k|^{1+2\alpha}}{\Gamma(1+\alpha^{-1})} \left(-\frac{1+\alpha/2}{2(1+2\alpha)} \left(\frac{1}{2} + (1+\alpha/2) \right) + \frac{1}{2\alpha(1+2\alpha)} \frac{1}{1-2\alpha} \left(\frac{\alpha}{2}\right)^2 \right) \;.
\end{eqnarray}
Finally, combining Eq. (\ref{comb_b1_b2}) and Eq. (\ref{eq_b3_asympt}) together with the small $k$ expansion of ${\hat F_{\alpha,2}(k)}$ above (\ref{f2_small_k}), one sees that the term proportional to $|k|^{1+\alpha}$ actually cancels, yielding 
\begin{eqnarray}\label{f_small_k_app}
&& \hat F_\alpha(k) = 1 + c_\alpha |k|^{1+2\alpha} + {\cal O}(k^{1+3\alpha}) \;, \\
&& c_\alpha = \frac{1}{\Gamma(1+\alpha^{-1})} \frac{2^{-2(2+\alpha)} \pi^{3/2} \tan{(\alpha \pi/2)} }{\cos{(\alpha \pi)} (1+\alpha) \Gamma(-\alpha) \Gamma(5/2+\alpha)} \;.
\end{eqnarray}
This singular behavior of $\hat F_\alpha(k)$ for small $k$ (\ref{f_small_k_app}) yields the power law behavior of $F_\alpha(Y)$ for large $Y$
\begin{eqnarray}\label{expr_appendix}
&& F_\alpha(Y) \propto \frac{a_\alpha}{Y^{2(1+\alpha)}} \;, \; Y \gg 1 \;, \\
&& a_\alpha = - \frac{1}{\pi} \Gamma(2+2\alpha) \cos{(\alpha \pi)} c_\alpha = \frac{2^{-2(2+\alpha)} \sqrt{\pi} \Gamma(2+2\alpha) \tan{(\alpha \pi/2)}}{\Gamma(2+\alpha^{-1}) \Gamma(1-\alpha) \Gamma(\frac{5}{2} + \alpha)} \;.
\end{eqnarray}

\subsection{The case $1/2 < \alpha \leq 1$}

This case can be studied along the same line as above except that in that case, $1 + 2\alpha > 2$ and therefore one has to handle with care the analysis of terms which are proportional to $k^2$, while the coefficient proportional to $k^{1+2\alpha}$ has the same form (\ref{f_small_k_app}). We will not repeat the analysis and simply give the result. One finds that $\hat F_\alpha(k)$ behaves for small $k$ as
\begin{eqnarray}\label{expr_bc}
 \hat F_\alpha(k) = 1 - \frac{b_\alpha}{2} k^2 + c_\alpha k^{1+2\alpha} + {\cal O}(k^{\min{(3,1+3\alpha)}})
\;, \; b_{\alpha} = \frac{\alpha \Gamma(2-\alpha^{-1})}{12 \Gamma(1+\alpha^{-1})} \;.
\end{eqnarray}
The expression for $b_\alpha$ given above (\ref{expr_bc}) yields the expression for $\langle Y^2 \rangle$ given in the text in Eq. (\ref{area_variance}).  

\subsection{The case $1 < \alpha \leq 2$}

In this case one can again perform a similar analysis but in this case one has $1+2 \alpha > 3$. And therefore one has to handle carefully the term proportional to $|k|^3$. A quite lengthy calculation shows that this term actually vanishes for $\alpha > 1$, while the coefficients of the terms proportional to $k^2$ and $|k|^{1+2\alpha}$ are still given by the expressions above (\ref{expr_bc}). This yields again as above (\ref{expr_bc})
\begin{eqnarray}
  \hat F_\alpha(k) = 1 - \frac{b_\alpha}{2} k^2 + c_\alpha k^{1+2\alpha} + {\cal O}(k^5) \;.
\end{eqnarray}

\section{Explicit expression of $F_\alpha(Y)$ for $\alpha=1$}\label{appendix_elementary}

In this appendix, we give an explicit expression of $F_1(Y)$ for $\alpha=1$. The starting point of our analysis is the expression (\ref{expr_arctan}) given
in the text:
\begin{eqnarray}
&& F_1(Y) = \frac{1}{\pi}\frac{2}{1+4Y^2} \\
&&+ \frac{1}{\pi} \left( \frac{2(1-8Y^2)}{(1+4Y^2)(1+16Y^2)} + \frac{4}{(1+16Y^2)^{\frac{3}{2}}} {\rm Re} \left[{(1- 4 i Y)^{\frac{3}{2}} \arctan{\left((1+4 i Y)^{-\frac{1}{2}}\right)}}\right]
\right) \nonumber 
\end{eqnarray}
For $z$ a complex number, the following elementary relations are useful :
\begin{eqnarray}
&& \arctan{z} = \frac{1}{2i} \left(\log{(1+i z)} - \log{(1-iz)} \right) \;, \\
&& \log{(x+ i y)} = \log{(\sqrt{x^2+y^2})} + 2 i \arctan{\left(\frac{y}{x + \sqrt{x^2+y^2}} \right)} \;.
\end{eqnarray}
On the other hand has
\begin{eqnarray}\label{def_ab}
 \frac{1}{\sqrt{1+ 4 i Y}} = a + i b \; , \; && a = \frac{1}{(1+16Y^2)^{1/4}} \cos{\left(\frac{\theta}{2}\right)} \;, \\
&& b =  \frac{-1}{(1+16Y^2)^{1/4}} \sin{\left(\frac{\theta}{2}\right)} \;,  \nonumber \\
&& \theta = \arctan{(4 Y)} \;. \nonumber 
\end{eqnarray}
Defining $\lambda$ and $\mu$ as
\begin{eqnarray}\label{def_lm}
\lambda &=& \arctan{\left[\frac{a^2+b^2-1+\sqrt{a^2+(1-b)^2}\sqrt{a^2+(1+b)^2}}{2a} \right]} \;, \\
 \mu &=& -\frac{1}{4} \log{\left[\frac{(1-b)^2+a^2}{(1+b)^2+a^2} \right]} \;, \nonumber 
\end{eqnarray}
in terms of $a, b$ defined above (\ref{def_ab}), one obtains finally (after straightforward algebra)
\begin{eqnarray}\label{elem}
 F_1(Y) = \frac{1}{\pi}\frac{2}{1+4Y^2} \\
+ \frac{1}{\pi} \left( \frac{2(1-8Y^2)}{(1+4Y^2)(1+16Y^2)} + \frac{4}{(1+ 16Y^2)^{3/4}} \left( \lambda \cos{\left(\frac{3\theta}{2} \right)} +   \mu \sin{\left(\frac{3\theta}{2} \right)} \right) \right) \;. \nonumber
\end{eqnarray}


\newpage

\section*{References}


\begin{thebibliography}{100}

\bibitem{chandrasekhar}
S. Chandrasekhar, Rev. Mod. Phys. {\bf 15}, 1 (1943). 


\bibitem{feller}
W. Feller, {\it An introduction to Probability Theory and its Applications}, (Wiley), New York (1968).

\bibitem{hughes}
B. Hughes, {\it Random walks and random environments}, (Clarendon
Press), Oxford (1968).

\bibitem{koshland}
D.E. Koshland, {\it Bacterial Chemotaxis as a Model Behavioral
  System}, (Raven), New York (1980). 

\bibitem{asmussen}
S. Asmussen, {\it Applied Probability and Queues}, (Springer), New
York (2003); M.J. Kearney, J. Phys. A {\bf 37}, 8421 (2004). 


\bibitem{satya_functionals}
S.N. Majumdar, {\it Brownian functionals in Physics and Computer Science}, 
Current Science {\bf 89}, 2076 (2005); {\it Universal First-passage Properties of Discrete-time Random Walks and L\'evy Flights on a Line: Statistics of the Global Maximum and Records}, Leuven Lectures FPSP-XII (2009), preprint arXiv:0912.2586 (to appear in Physica A). 

\bibitem{williams}
R.J. Williams, {\it Introduction to the Mathematics of Finance},
(AMS), (2006); M. Yor, {\it Exponential Functionals of Brownian Motion
and Related Topics}, (Springer), Berlin (2000). 



\bibitem{rivasseau}
J. de Coninck, F. Dunlop, V. Rivasseau, Commun. Math. Phys. {\bf 121}, 401 (1989).

\bibitem{satya_airy}
S.~N.~Majumdar, A.~Comtet, Phys. Rev. Lett. {\bf 92}, 225501 (2004);
J. Stat. Phys. {\bf 119}, 777 (2005).

\bibitem{kearney}
M. J. Kearney, S.N. Majumdar, J. Phys. A: Math. Gen. {\bf 38}, 4097 (2005); M. J. Kearney, S.N. Majumdar, R.J. Martin, J. Phys. A: Math. Theor. {\bf 40}, F863 (2007).

\bibitem{schehr_airy}
G. Schehr, S.N. Majumdar, Phys. Rev. E {\bf 73}, 056103 (2006).

\bibitem{welinder}
P. Welinder, G. Pruessner, K. Christensen, New J. of Phys. {\bf 9}, 149 (2007). 


\bibitem{janson_review}
S. Janson, Proba. Survey {\bf 4}, 80 (2007).

\bibitem{rajabpour}
M. Rajabpour, J. Phys. A : Math. Theor. {\bf 42}, 485205 (2009). 

\bibitem{rambeau_airy}
J. Rambeau, G. Schehr,  J. Stat. Mech., P09004 (2009).


\bibitem{dhar}
D. Dhar, R. Ramaswamy, Phys. Rev. Lett. {\bf 63}, 1659 (1989).  

\bibitem{pld}
P. Le Doussal, K. J. Wiese, Phys. Rev. E {\bf 79}, 051105 (2009). 




\bibitem{waclaw}
B. Waclaw, J. Sopik, W. Janke, H. Meyer-Ortmanns, Phys. Rev. Lett. {\bf 103}, 080602 (2009); J. Stat. Mech. P10021 (2009).

\bibitem{satya_condensation}
M. R. Evans, T. Hanney, S. N. Majumdar, Phys. Rev. Lett. {\bf 97}, 010602 (2006) 

\bibitem{evans_review}
M. R. Evans, T. Hanney, J. Phys. A: Math. Gen. {\bf 38}, R195, (2005).

\bibitem{godreche_review}
C. Godr\`eche, Lect. Notes Phys. {\bf 716} 261 (2007), arXiv:cond-mat/0604276. 


\bibitem{satya_review_condensation}
S. N. Majumdar, Les Houches lecture notes for the summer school {\it Exact Methods in Low-dimensional Statistical Physics and Quantum Computing}, (2008), preprint arXiv:0904.4097. 



\bibitem{knight}
F.B. Knight, {\it Hommage \`a P.A. Meyer et J. Neveu}, {Ast\'erisques}, 171 (1996); L. Chaumont, D.G. Hobson, M. Yor, S\'em. de Prob. XXXV, 334, (2001). 

\bibitem{bertoin}
J. Bertoin, {\it L\'evy processes}, Camb. Univ. Press., Melbourne, NY, (1996). 

\bibitem{rap}
For a short review see T.W. Burkhardt, J. Stat. Mech. P07004 (2007).

\bibitem{kawasaki}
K. Kawasaki, Phys. Rev. {\bf 145}, 224 (1966). 

\bibitem{bertin_review}
M. Clusel, E. Bertin, Int. J. Mod. Phys. B {\bf 22}, 3311 (2008). 









\end{thebibliography}
\end{document}